\documentclass[twocolumn,preprint]{aastex63}
\usepackage{amsmath,amstext}
\usepackage[T1]{fontenc}
\usepackage{apjfonts} 
\usepackage{natbib}
\usepackage{longtable}

\newcommand{\hst}{\textit{HST}}
\newcommand{\HST}{\textit{HST}}

\newcommand{\Paa}{\hbox{{\rm Pa}$\alpha$}}
\newcommand{\Pab}{\hbox{{\rm Pa}$\beta$}}
\newcommand{\Ha}{\hbox{{\rm H}$\alpha$}}
\newcommand{\Hb}{\hbox{{\rm H}$\beta$}}

\begin{document}

\title{\large \bf CLEAR: Paschen-$\boldsymbol{\beta}$ Star-Formation Rates and Dust Attenuation of Low-Redshift Galaxies}

\author[0000-0001-7151-009X]{Nikko J. Cleri}
\affil{Department of Physics, University of Connecticut, Storrs, CT 06269, USA}
\affiliation{Department of Physics and Astronomy, Texas A\&M University, College
Station, TX, 77843-4242 USA}
\affiliation{George P.\ and Cynthia Woods Mitchell Institute for
Fundamental Physics and Astronomy, Texas A\&M University, College
Station, TX, 77843-4242 USA}

\author[0000-0002-1410-0470]{Jonathan R. Trump}
\affil{Department of Physics, University of Connecticut, Storrs, CT 06269, USA}

\author[0000-0001-8534-7502]{Bren E. Backhaus}
\affil{Department of Physics, University of Connecticut, Storrs, CT 06269, USA}

\author[0000-0003-1665-2073]{Ivelina Momcheva}
\affil{Space Telescope Science Institute, 3700 San Martin Drive,
Baltimore, MD, 21218 USA}

\author[0000-0001-7503-8482]{Casey Papovich}
\affiliation{Department of Physics and Astronomy, Texas A\&M University, College
Station, TX, 77843-4242 USA}
\affiliation{George P.\ and Cynthia Woods Mitchell Institute for
Fundamental Physics and Astronomy, Texas A\&M University, College
Station, TX, 77843-4242 USA}

\author[0000-0002-6386-7299]{Raymond Simons}
\affil{Space Telescope Science Institute, 3700 San Martin Drive,
Baltimore, MD, 21218 USA}

\author[0000-0001-6065-7483]{Benjamin Weiner}
\affil{MMT/Steward Observatory, 933 N. Cherry St., University of Arizona, Tucson,
AZ 85721, USA}

\author[0000-0001-8489-2349]{Vicente Estrada-Carpenter}
\affiliation{Department of Physics and Astronomy, Texas A\&M University, College
Station, TX, 77843-4242 USA}
\affiliation{George P.\ and Cynthia Woods Mitchell Institute for
Fundamental Physics and Astronomy, Texas A\&M University, College
Station, TX, 77843-4242 USA}

\author[0000-0001-8519-1130]{Steven L. Finkelstein}
\affil{Department of Astronomy, The University of Texas, Austin, Texas, 78712 USA} 

\author{Mauro Giavalisco}
\affil{Department of Astronomy, University of Massachusetts,
Amherst, MA, 01003 USA} 

\author[0000-0001-7673-2257]{Zhiyuan Ji}
\affil{Department of Astronomy, University of Massachusetts,
Amherst, MA, 01003 USA}

\author[0000-0003-1187-4240]{Intae Jung}
\affil{Department of Physics, The Catholic University of America, Washington, DC 20064, USA}
\affil{Astrophysics Science Division, Goddard Space Flight Center, Greenbelt, MD 20771, USA}

\author[0000-0002-7547-3385]{Jasleen Matharu}
\affiliation{Department of Physics and Astronomy, Texas A\&M University, College
Station, TX, 77843-4242 USA}
\affiliation{George P.\ and Cynthia Woods Mitchell Institute for
Fundamental Physics and Astronomy, Texas A\&M University, College
Station, TX, 77843-4242 USA}

\author[0000-0002-9883-1413]{Felix Martinez III}
\affiliation{Department of Physics and Astronomy, Texas A\&M University, College
Station, TX, 77843-4242 USA}
\affiliation{George P.\ and Cynthia Woods Mitchell Institute for
Fundamental Physics and Astronomy, Texas A\&M University, College
Station, TX, 77843-4242 USA}

\author[0000-0003-1055-1888]{Megan R. Sturm}
\affil{Department of Physics, University of Connecticut, Storrs, CT 06269, USA}

\begin{abstract}
We use \Pab\ (1282~nm) observations from the Hubble Space Telescope ($\HST$) G141 grism to study the star-formation and dust attenuation properties of a sample of 29 low-redshift ($z < 0.287$) galaxies in the CANDELS Ly$\alpha$ Emission at Reionization (CLEAR) survey. We first compare the nebular attenuation from $\Pab/\Ha$ with the stellar attenuation inferred from the spectral energy distribution, finding that the galaxies in our sample are consistent with an average ratio of the continuum attenuation to the nebular gas of 0.44, but with a large amount of excess scatter beyond the observational uncertainties. Much of this scatter is linked to a large variation between the nebular dust attenuation as measured by (space-based) $\Pab$ to (ground-based) $\Ha$ to that from (ground-based) $\Ha/\Hb$.  This implies there are important differences between attenuation measured from grism-based / wide-aperture $\Pab$ fluxes and the ground-based / slit-measured Balmer decrement.  We next compare star-formation rates (SFRs) from $\Pab$ to those from dust-corrected UV.  We  perform a survival analysis to infer a census of \Pab\ emission implied by both detections and non-detections. We find evidence that galaxies with lower stellar mass have more scatter in their ratio of \Pab\ to attenuation-corrected UV SFRs. When considering our \Pab\ detection limits, this observation supports the idea that lower mass galaxies experience ``burstier'' star-formation histories.  Together, these results show that \Pab\ is a valuable tracer of a galaxy's SFR, probing different timescales of star-formation and potentially revealing star-formation that is otherwise missed by UV and optical tracers.
\end{abstract}

\section{Introduction}\label{sec:intro}

Star-formation rates (SFRs) are a critical quantity in the understanding of galaxy evolution. There exist several different methods of estimating SFR in a given galaxy, the most direct of which is counting identifiable stars of a specific age \citep{kenn12}. With current instrumentation, this method of star counting is limited to the most immediate of Milky Way satellites. In more distant galaxies, the primary methods of measuring SFRs are using continuum and emission-line tracers \citep[e.g.,][]{kenn12}.

Near-ultraviolet (UV) continuum observations of a galaxy measure the photospheric emission of massive young stars, and so the UV continuum acts as a direct tracer of recent star-formation, timescales of order hundreds of Myr \citep{kenn12,redd12}. However, UV continuum observations are highly sensitive to attenuation by interstellar dust. In principle, this attenuation can be corrected using the UV slope $\beta$, but in practice the unknown intrinsic UV slope and differences in the UV shape of different attenuation laws complicate this approach \citep[e.g.,][]{salim20}. Another approach is to add the reprocessed IR emission to the observed UV continuum for a "ladder" SFR \citep{wuyt11,whit14}, but this is similarly complicated by UV optical depth effects and/or potential anisotropy of the IR emission \citep{kenn12,barr19}.

Optical and near-infrared (IR) emission lines from ionized gas around massive stars are also widely used as SFR indicators. These emission lines receive peak contribution from stars of mass 30-40 $\text{M}_\odot$, and as such are tracers of stars with lifetimes of 3-10 Myr. Recombination lines of hydrogen are especially useful to trace star-formation since they are insensitive to metallicity and relatively insensitive to gas temperature and density \citep{oste89}. Since the continuum and emission-line tracers correspond to star-formation on different timescales, their ratio can be used to measure the burstiness of the star-formation \citep[e.g.,][]{guo16,weis12}. Still, optical emission lines are susceptible to dust attenuation. For example, \Ha\ flux is reduced by a factor of $\sim$2 at a modest attenuation of $A_V = 1$, and reduced by a factor of $\sim$10 in a dusty galaxy with $A_V = 3$. The Balmer decrement ($\Ha/\Hb$) can be used to correct for the attenuation, but this correction is inaccurate in regions of high optical depth to Balmer emission and can be inaccurate if the emission and/or attenuation scales are smaller than the spatial resolution \citep{kenn12}. Uncertainties in correcting for dust attenuation fundamentally limit measurements of star-formation burstiness from the UV continuum and optical emission-line SFR tracers \citep{brou19}.

Near-IR recombination lines of hydrogen offer a solution to the problem of dust attenuation in measuring SFR. Just like the more commonly used Balmer series, the Paschen lines of hydrogen are highly sensitive to the ionizing ($E>13.6$~eV) radiation of OB stars formed within the last 10 Myr, while remaining relatively insensitive to nuisance parameters like the temperature and density of the star-forming gas \citep{oste89}. Unlike the optical Balmer lines, the near-IR Paschen lines are far less affected by interstellar dust extinction, so the Paschen lines can reveal otherwise hidden star-forming regions that are shrouded in gas and dust that is optically thick to Balmer emission.

In previous work, \Pab\ and \Ha\ have been studied in a 2 galaxy sample \citep{kess20}. Previous studies have also used the \Paa\ (18750\AA) emission line to calibrate mid-IR SFR indicators in nearby starburst and luminous IR galaxies \citep{alon06,calz07} and in rare lensed galaxies at higher redshift \citep{papo09,fink11,ship16}.

In this work we study \Pab\ (1282 nm), the $n = 5 \rightarrow 3$ hydrogen recombination line, as an SFR indicator. We use \Pab\ fluxes measured from near-IR spectroscopy from the \textit{HST}/WFC3 grisms taken as part of the 3D-HST \citep{momc16} and CLEAR \citep{simo20}
surveys, as described in Section \ref{sec:data}. Section \ref{sec:dust} discusses the viability of \Pab/\Ha\ as an attenuation indicator compared to the Balmer decrement and V-band continuum attenuation. In Section \ref{sec:sfr}, we compare \Pab\ to other SFR indicators, demonstrating that \Pab\ includes star-formation missed by UV and optical tracers and showing evidence for burstier star formation at low mass. We summarize our results and discuss future applications with the James Webb Space Telescope (\textit{JWST}) in Section \ref{sec:conclusions}.

Throughout this work, we assume a WMAP9 cosmology with $\Omega_{m,0} = 0.287$, $\Omega_{\Lambda,0} = 0.713$, and $H_0 = 69.3 \text{ km s}^{-1}\text{ Mpc}^{-1} $ \citep{hins13}. We also assume intrinsic line ratios of $\Ha/\Hb=2.86$ and $\Ha/\Pab=17.6$, corresponding to Case B recombination at a temperature of $T=10^4$~K and a density of $n_e = 10^4$~cm$^{-3}$ \citep{oste89}. 

\section{Data}\label{sec:data}

Our data come from the CLEAR survey (a Cycle 23 $\hst$ program, PI: Papovich), which consists of deep (12-orbit depth) \HST/WFC3 G102 slitless grism spectroscopy covering $0.8-1.2$~\micron\ within 12 fields split between the GOODS-North (GN) and GOODS-South (GS) extragalactic survey fields (\citealt{estr19, simo20}). The CLEAR pointings overlap with the larger 3D-HST survey area \citep{momc16}, which provides slitless G141 grism spectra of 2-orbit depth and a spectral wavelength range of 1.1-1.65~\micron.

\subsection{CLEAR Parent Sample G102 and G141 Spectroscopy, Redshifts and Line Fluxes}

The \texttt{grizli} (grism redshift and line analysis) pipeline \citep{bram19} serves as the primary method of data reduction for the CLEAR dataset. In contrast to traditional methods of extracting one-dimensional (1D) spectra from slit observations, \texttt{grizli} directly fits the two-dimensional (2D) spectra with model spectra convolved to the galaxy image and for multiple position angles of grism observations. This process yields complete and uniform characterization of the suite of spectral line features of all objects observed in each of the G102 and G141 grisms. The most relevant of these spectral properties for our analysis are redshifts, line fluxes, and emission-line maps. The \Pab\ line is not included in the \texttt{grizli} fits by default, but was included in the CLEAR reductions for this work.

Our parent sample represents all CLEAR galaxies within the redshift range for detectable \Pab\ in the G141 spectrum ($z<0.287$). The CLEAR spectral extractions are limited to galaxies with $m_{\rm F105W} < 25$.

\subsection{Sample Selection}\label{subsec:sample}

We select a sample of \Pab-emitting galaxies from the CLEAR parent catalog using the following steps:
\begin{itemize}
    \item Require $z<0.287$, such that \Pab\ is within the observed-frame spectral range and blueward of the G141 sensitivity decline at 1.65~$\mu$m.
    \begin{itemize}
        \item Primary sample: A \Pab\ signal-to-noise ratio (SNR) of ${\rm SNR}>3$. (20 objects)
        \item Secondary sample: A marginal \Pab\ ${\rm SNR}>1$ and a reliable spectroscopic redshift from either ground-based optical spectroscopy or from \Ha\ emission in the G102 spectrum. (9 objects)
    \end{itemize}
\end{itemize}

The primary sample ensures reliable \Pab-detections, and the secondary sample ensures reliable redshifts from other (brighter) emission lines, while remaining as inclusive as possible to non-spurious \Pab\ emission. The combined sample is constructed to include all non-spurious \Pab-detections, even when the \Pab\ SNR is only marginal. For clarity in all of the following figures, objects in our primary sample will be plotted with larger symbol sizes than those in our secondary sample.

Our total (primary+secondary) sample includes 29 \Pab-emitting galaxies: 20 from the primary sample and 9 from the secondary sample. This total sample comprises approximately $19\%$ of all CLEAR galaxies in this redshift range. The median \Pab\ SNR of the total sample is 3.9 with the median SNR of the >3$\sigma$ primary sample detections of 5.1. The grism redshifts and \Pab\ line fluxes for the objects in our sample are taken from the CLEAR release v2.1.0 \citep{simo20}. We note that 17 of the objects in our sample have matching spectroscopic redshifts from ground-based programs (as compiled in the 3D-HST catalog), and another 6 have a redshift from well-detected \Ha\ in the G102 grism.

\begin{figure}[t]
\epsscale{1.1}
\plotone{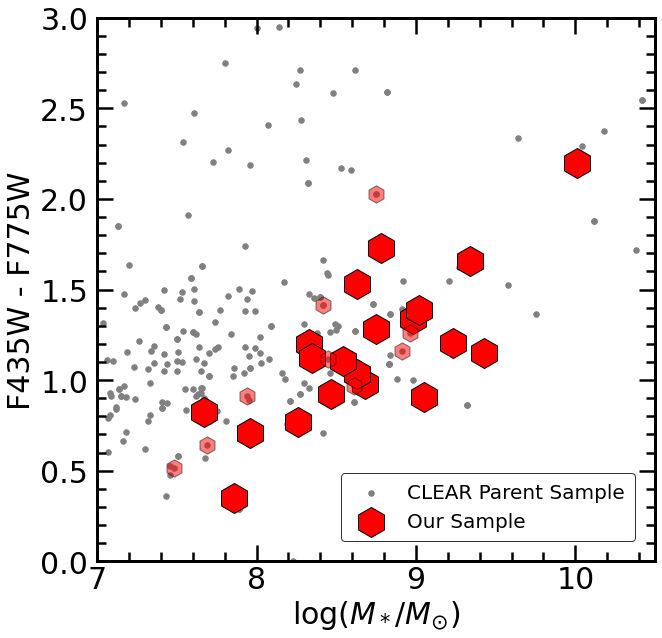}
\caption{The relation between F435W-F775W color and stellar mass for galaxies of redshift $z<0.287$. Our sample is shown as red stars, with the rest of the CLEAR galaxies in this redshift range shown as gray points. Larger symbols represent objects in our primary sample with \Pab\ SNR>3, and smaller symbols represent objects in our secondary sample with reliable redshifts and marginal \Pab\ 1<SNR<3. The galaxies in our sample are broadly representative of the population of $z<0.287$ star-forming galaxies with $\log(M_*/M_\odot) \gtrsim 8$. The sample of \Pab-detected galaxies also includes a few red galaxies that are likely dust-obscured (see Section \ref{sec:dust}).
\label{fig:colormag}}
\end{figure}

Figure \ref{fig:colormag} shows the color-mass relation for the galaxies in our \Pab-detected sample (in red) with respect to all CLEAR galaxies in the redshift range $z < 0.287$ (in gray). The F435W and F775W magnitudes and stellar masses are taken from the CANDELS/SHARDS multiwavelength catalog \citep{barr19}. Our galaxy sample is broadly representative of the larger galaxy population in CLEAR, with some preference for blue galaxies with $\log(M_*/M_\odot) \gtrsim 8$.

\begin{figure}[t]
\epsscale{1.1}
\plotone{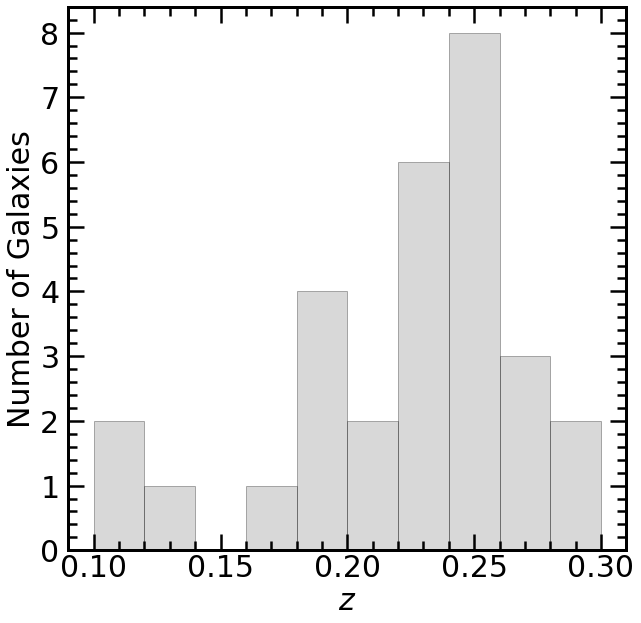}
\caption{Histogram of our sample of \Pab-selected objects binned by grism redshift. The G141 grism wavelength range limits the detection of \Pab\ to $z<0.287$. Our sample has a redshift range of $0.11 < z < 0.28$, with a median grism redshift of 0.23.
\label{fig:zhist}}
\end{figure}

\begin{figure*}[t]
\epsscale{1.15}
\plotone{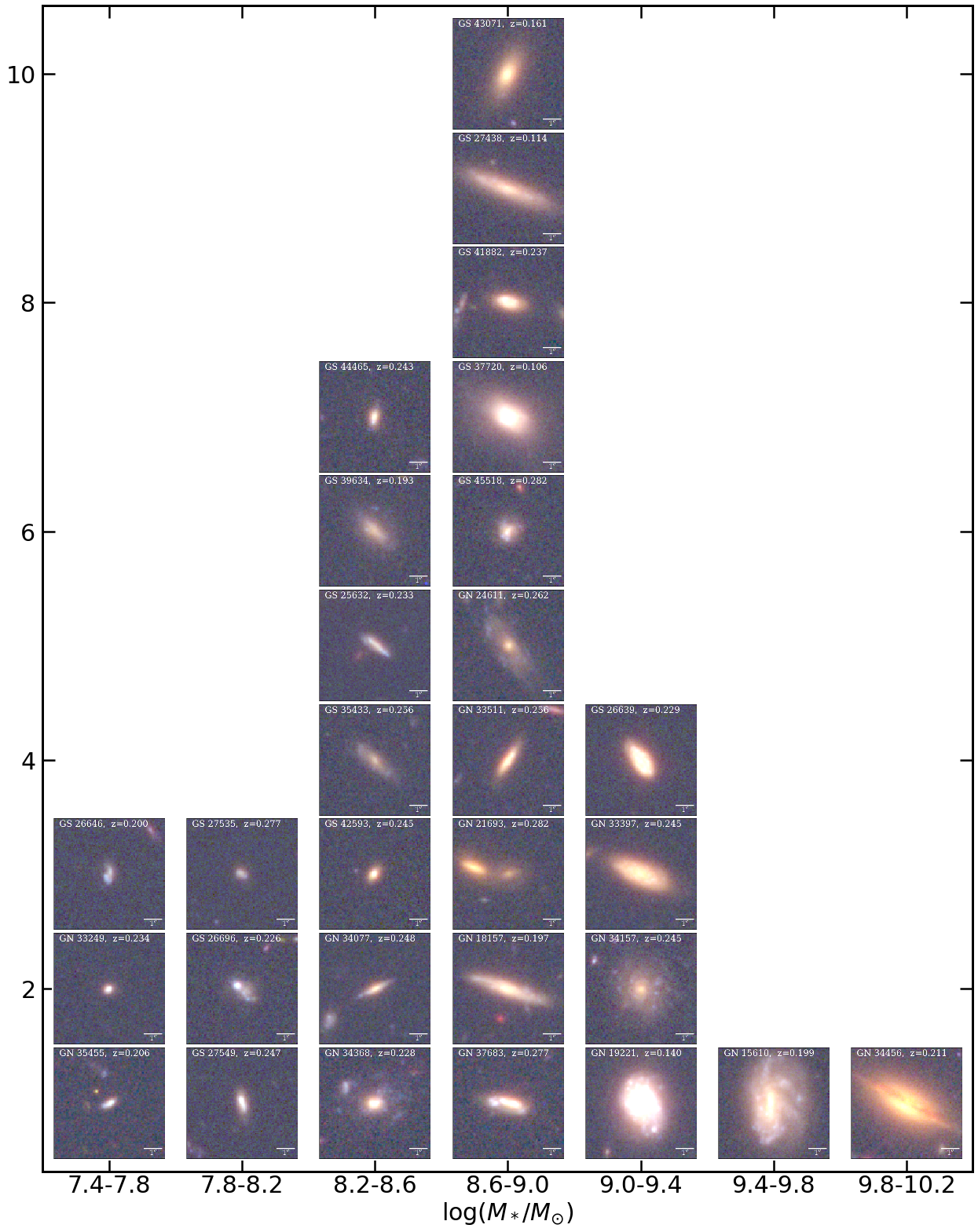}
\caption{RGB images of each galaxy in the sample, binned by stellar mass. Our sample includes a diverse range of galaxy morphologies, from compact, low-stellar mass objects to extended higher-mass spirals and ellipticals.
\label{fig:images}}
\end{figure*}

Figure \ref{fig:zhist} shows the redshift distribution of the galaxies in our sample. The redshift range of objects in our sample is $0.11 < z < 0.28$ with a median redshift $z = 0.23$. Figure \ref{fig:images} shows RGB images ($i$, $Y$, and $H$ band) of all 29 \Pab-detected galaxies in our primary and secondary samples, binned by stellar mass. Our sample includes a broad range of galaxy morphologies, including high-mass extended disks, bright compact sources, and diffuse irregular objects.

\begin{figure*}[t]
\epsscale{1.2}
\plotone{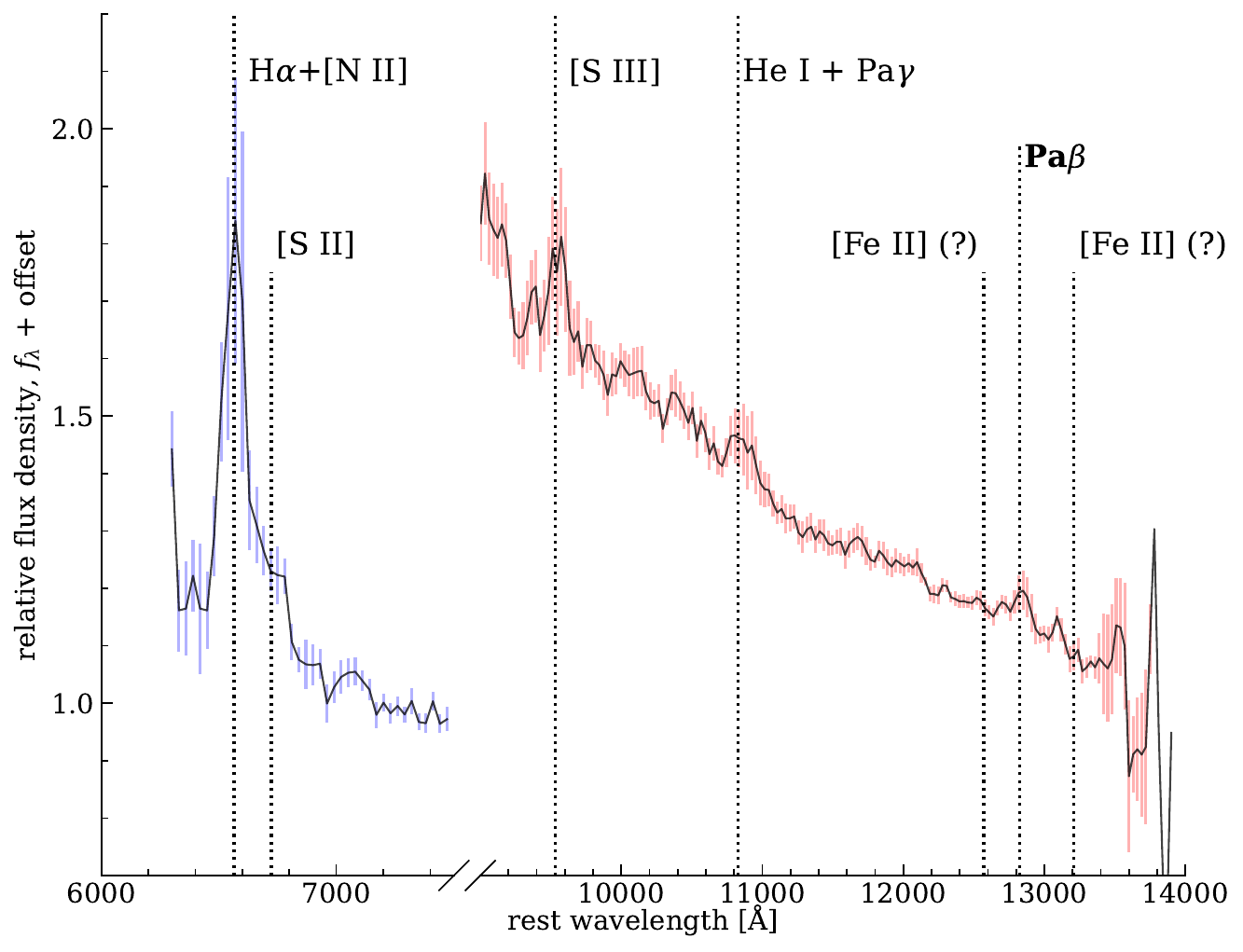}
\caption{Stacked spectra of objects in the primary and secondary sample with \Ha\ signal to noise ratio greater than 2. We also show several other emission-line features visible in the G102 (blue) and G141 (red) spectral coverage. We observe stacked fluxes corresponding to the relatively faint intrinsic ratio of $\Ha/\Pab = 17.6$ assuming Case B recombination as described in Section \ref{sec:intro}. We show an example spectrum for a single object in Figure \ref{fig:spectra}. 
\label{fig:stack}}
\end{figure*}

Figure \ref{fig:stack} shows the stacked one dimensional spectrum for all objects in the sample with \Ha\ SNR > 2. We also show a suite of several other emission features in the near-IR, including relatively strong emission from doubly ionized sulfur (\ion{S}{3}). The stacked emission features are visibly consistent with the intrinsic \Pab/\Ha\ ratio of 1/17.6. We note that the stacked emission-line fluxes and flux ratios do not affect the main conclusions of the paper (see Figure \ref{fig:survival}), which rely on the scatter of measurements between individual galaxies.

\begin{figure*}[t]
\epsscale{1.2}
\plotone{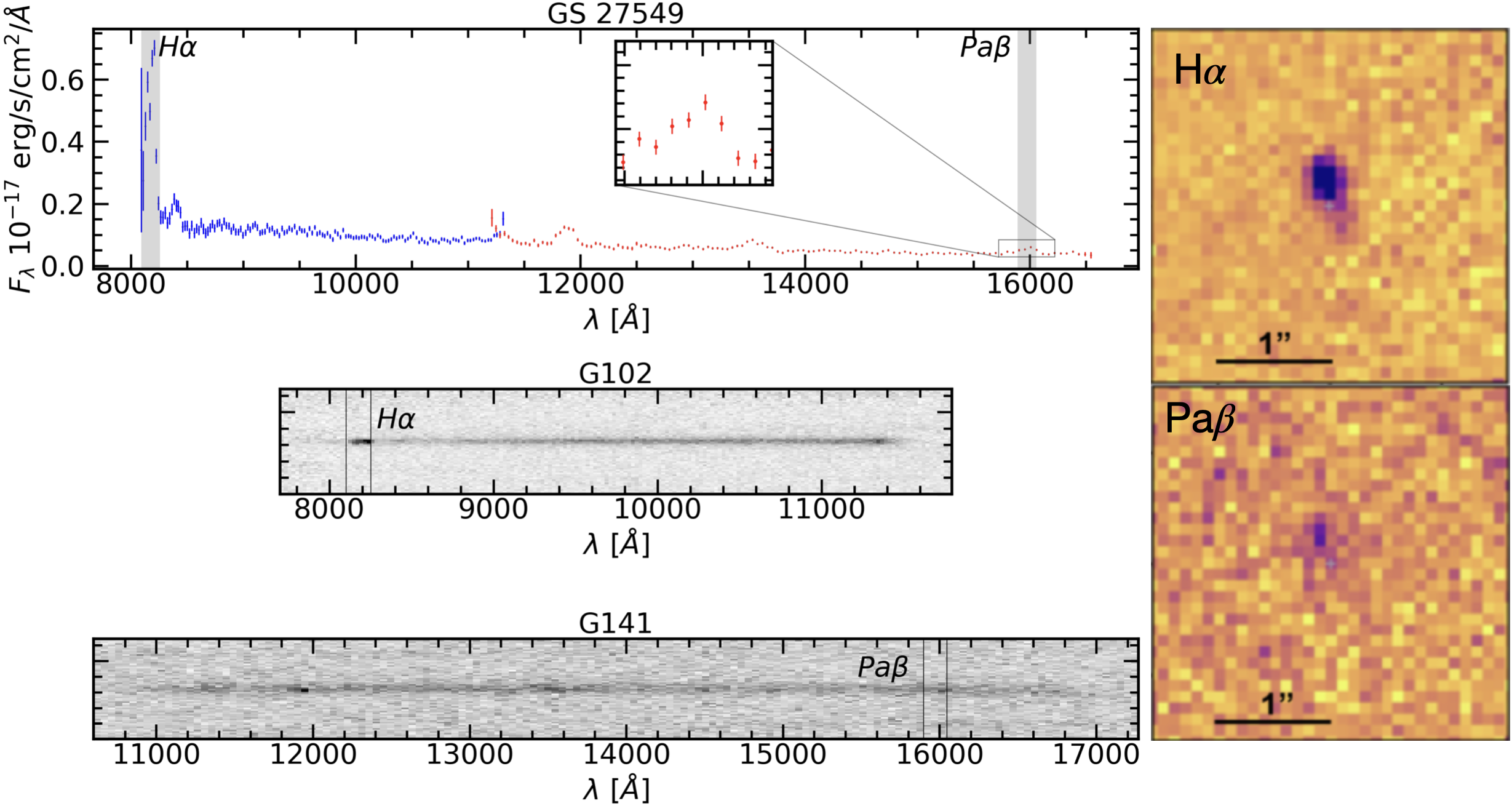}
\caption{\textit{Top left}: Observed-frame one-dimensional (1D) spectrum for a galaxy in our sample. The G102 (blue) and G141 (red) spectra show the median points from all exposures for this object. The inset shows the region around the \Pab\ line. The gray shaded regions show the \Pab\ and \Ha\ lines.  \Ha\ is available in the G102 in a small redshift window where \Pab\ is simultaneously available in the G141 ($0.22<z<0.287$). \textit{Bottom left}: Observed-frame two-dimensional (2D) spectra for the same galaxy in our sample. We indicate both \Pab\ and \Ha\ (where available) by annotated regions outlined in black lines. \textit{Right}: Spatially-Resolved emission-line maps of \Ha\ (top) and \Pab\ (bottom) for the same object from the \texttt{grizli} extractions. \Pab\ is a relatively weak line, with the intrinsic ratio of \Ha/\Pab\ $\approx 17.6$ \citep{oste89}. 
\label{fig:spectra}}
\end{figure*}

Figure \ref{fig:spectra} shows one-dimensional (1D) and two-dimensional (2D) spectra for a galaxy in our \Pab-detected sample. We show one (GS 27549) object with strong enough \Pab\ SNR to extract spatially-resolved attenuation maps. The points of the 1D spectra show only the median points of the each individual grism exposures for this object. We note that the intrinsic \Pab/\Ha\ ratio is a relatively weak 1/17.6. The spatially-resolved line maps of \Ha\ and \Pab\ show nebular emission visually consistent with the intrinsic \Pab/\Ha\ ratio \citep{oste89}.

Our CLEAR \Pab-detected sample may include a bias to high \Pab\ fluxes that are not representative of the full galaxy population, especially considering the line flux limit of $1.5 \times 10^{-17}$ erg s$^{-1}$ cm$^{-2}$ for the G141 grism observations \citep{momc16}. We use the \Pab\ flux limits, given as the line flux uncertainty from \texttt{grizli} when it tries to fit a \Pab\ line, for the rest of the 152 CLEAR galaxies in the same $z<0.287$ redshift range. We consider this sample of galaxies with \Pab\ flux limits in a survival analysis in Section \ref{subsec:burstiness} that considers the relationship between \Pab\ and UV SFRs for both \Pab\ detections and non-detections.


\subsection{Photometry and Derived Quantities}\label{subsec:derived}

We take stellar masses for objects in our sample from the 3D-HST catalog \citep{skel14}, derived from the CANDELS photometry \citep{grog11, koek11}. The stellar masses are calculated with FAST \citep{krie09}, using a \cite{bruz03} stellar population synthesis model library, a \cite{chab03} IMF, solar metallicity, and assuming exponentially declining star-formation histories. The stellar masses of our $z<0.287$ galaxies are generally robust to these assumptions because the peak of the stellar emission is well-constrained by the high-quality CANDELS near-IR imaging. We additionally use the $V$-band attenuation ($A_V$) measured from the same FAST fit to the spectral energy distribution.

We use UV continuum SFRs from the catalog of \cite{barr19}, which supplements the CANDELS multiwavelength data with SHARDS photometry \citep{pere13} in GOODS-N. Attenuation-corrected UV SFRs are calculated using the \cite{kenn98} calibration with a dust attenuation correction \citep{barr19}:
\begin{equation}\label{eq:sfruvcorr}
    SFR_{UV}^{corr} [M_{\odot}yr^{-1}] = (1.09 \times 10^{-10}) (10^{0.4 A_{280}}) (3.3~L_{280}/L_{\odot})
\end{equation}
$L_{280}$ and $A_{280}$ are the UV luminosity and dust attenuation at rest-frame $\lambda = 280$~nm, respectively. The UV luminosity $L_{280} \equiv \nu L_{\nu}(280~\text{nm})$ is calculated from EAZY with a best-fit spectral energy distribution \citep{bram08,wuyt11}. The UV attenuation is inferred iteratively, measured from the best-fit SED while ensuring consistency with the IR (non)detection and the star-formation mass sequence (see Appendix D of \citealt{barr19}).  The shortest wavelength filter in the \cite{barr19} catalog is the $U$ band covering observed-frame $\lambda \approx 320$ nm at its bluest end. This iterative approach is designed to produce attenuation-corrected UV SFRs that are robust to poorly measured UV photometry, which is especially useful given the limited rest-frame UV coverage of the low-redshift galaxies in our sample.

The conversion factor for the UV luminosity is derived in \cite{bell05}. The attenuation-corrected UV SFR calibrations are metallicity-dependent, with a systematic uncertainty of 0.05 dex from Solar to 20\% Solar based on the \cite{bruz03} models.

The peak timescale probed by the near-UV-derived SFRs at 2800\AA\ is of order hundreds of Myr. \citep{redd12}. SFRs derived from bluer luminosities probe slightly shorter timescales, of order 10-100 Myr for SFRs from $L_{1500}$ \citep{redd12}. 

UV + IR ``ladder'' SFRs are calculated for objects with mid/far-IR detections following \citet{wuyt11}:
\begin{equation}\label{eq:sfrladder}
\text{SFR}_{UV+IR} [M_\odot yr^{-1}] = 1.09 \times 10^{-10}(L_{IR} + L_{280})L_{\odot}    
\end{equation}

The relative scale of the UV and IR contribution is based on local universe calibrations \citep{kenn12}, and the overall scale assumes a \cite{chab03} initial mass function. The UV + IR ladder SFR ultimately measures the UV continuum emission that is not attenuated by dust plus the reprocessed IR continuum emission from the UV which is absorbed by the dust. For objects which do not have IR detections, the ladder SFRs are defined to be equal to the attenuation-corrected UV SFRs of Eq \ref{eq:sfruvcorr} \citep{barr19}.

\begin{figure*}[t]
\epsscale{1.15}
\plotone{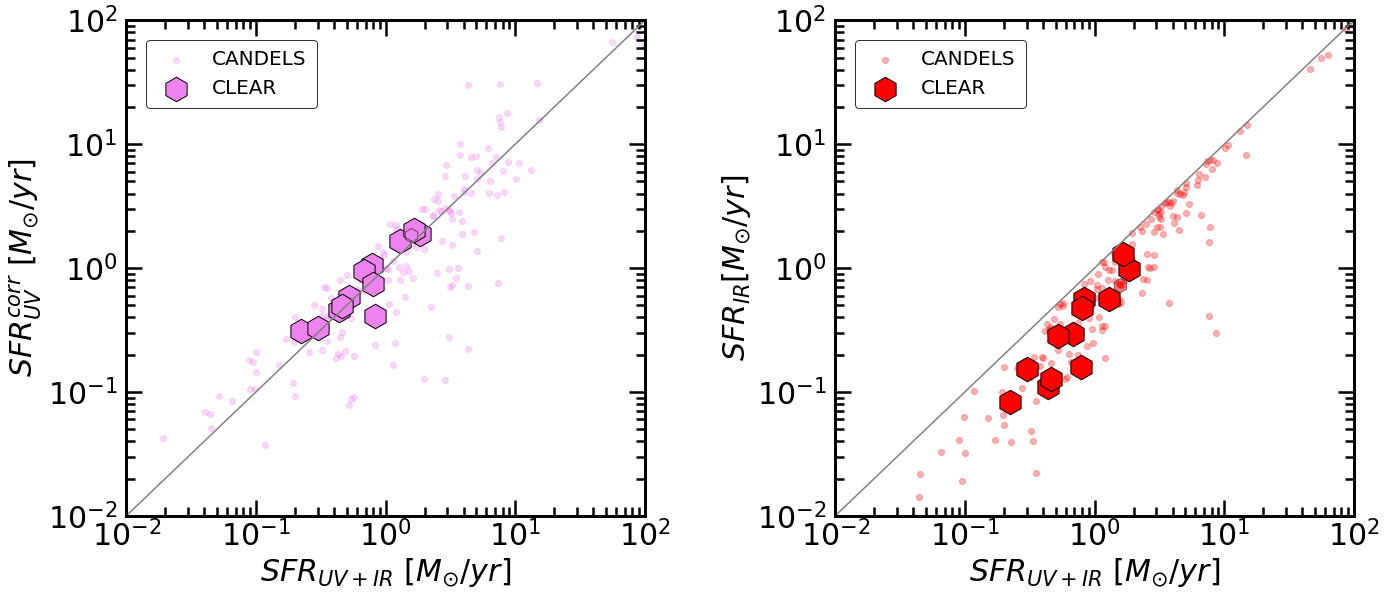}
\caption{\textit{Left}: The relation between attenuation-corrected UV continuum SFRs and UV+IR ``ladder'' SFRs for IR-detected galaxies in our CLEAR sample (large hexagons) and the \cite{barr19} CANDELS galaxies (small circles) for the $z<0.287$ regime. \textit{Right}: The relation between IR SFRs and UV+IR ``ladder'' SFRs for the same objects. We exclude the IR non-detections because these objects are defined to have equivalent attenuation-corrected UV and ``ladder'' SFRs. The attenuation-corrected UV SFRs more accurately models the true SFR than the IR SFR, which underestimates the total SFR by missing star-formation only visible in the UV.}
\label{fig:sfrcomparison}
\end{figure*}

In Figure \ref{fig:sfrcomparison}, we compare the attenuation-corrected UV SFRs and the UV + IR ladder SFRs for CLEAR galaxies with mid-IR detections. For the CLEAR sample, the attenuation-corrected UV and UV+IR SFR indicators agree with each other with a median absolute deviation of $\sim$0.09, similar to the scatter reported in \citet{barr19}. This indicates that the attenuation-corrected UV SFRs are likely to be reliable for our sample of galaxies.

Our galaxies have morphology measurements from \citet{van12}. We use effective (50\% light) radii and \citet{sers68} indices for galaxies with ``good'' GALFIT \citep{peng10} fits with flag = 0 (see \citealt{van12} for details). From these effective radii and S\'ersic indices, we calculate the central density within 1 kpc, $\Sigma_{1kpc}$.

To correct for dust attenuation of \Pab, \Ha, and \Hb, we assume a \citep{calz00} attenuation model. We use \cite{calz00} over other attenuation models of the Milky Way or the Small Magellanic Cloud \citep{fitz99,gord03} to maintain consistency with the attenuation-corrected UV SFRs from \cite{barr19}. The choice of attenuation model has little impact for this work since the attenuation models are very similar at optical and near-IR wavelengths (i.e., for the Balmer and Paschen lines). We use the \cite{calz00} attenuation model in accordance with previous studies showing a nebular-to-stellar attenuation ratio of $\approx$ 2 \citep{salim20,calz00}.

\subsection{Optical Spectra}\label{subsec:optical}

A subsample of 11 galaxies in GOODS-N match to publicly available optical spectra from the Team Keck Treasury Redshift Survey (TKRS, \citealt{wirt04}), from which we use \Ha\ and \Hb\ fluxes. The TKRS spectroscopic observations of GOODS-N were taken using DEIMOS on the Keck~II telescope, with the spectra extracted using the DEEP2 Redshift Survey Team pipeline \citep{newm13}. Disk-integrated Balmer line fluxes are estimated from TKRS spectra in a way that accounts for slit losses under the assumption that emission line equivalent widths are invariant across the stellar disk; see Section 2 of \cite{wein07} for details. We note that this assumption of invariant equivalent widths makes the TKRS measurements potentially susceptible to issues when measuring Balmer-line fluxes for large objects with significant color gradients. We discuss this further in Section \ref{subsec:pabha}.

\subsection{Linear Regression Methods and Significance of Fits}
Throughout this analysis, we use the \texttt{linmix} package in python \citep{kell07} to calculate our linear regression fits. For the remainder of this work, we consider a correlation between two quantities to be significant if the \texttt{linmix} mean best fit slope is 3$\sigma$ different from zero. The standard deviation $\sigma$ is the standard deviation of the \texttt{linmix} best-fit slopes.

\section{\Pab\ and Dust Attenuation} \label{sec:dust}

Because the ratios of the fluxes of recombination lines of hydrogen are relatively insensitive to metallicity, temperature, and density, their ratios can be used to estimate dust attenuation. The most commonly used emission-line indicator of dust attenuation is the Balmer decrement, $\Ha/\Hb$, which has an intrinsic ratio of 2.86 assuming Case B recombination with $T=10^4$~K and $n_e=10^4$~cm$^{-3}$ \citep{oste89}. This rest-frame optical ratio only works for modestly attenuated galaxies, since at $A_V \sim 1-2$ Balmer decrement attenuation measurements will entirely miss regions of the ISM that are optically thick to \Hb\ emission.

\subsection{\Pab\ and Nebular Attenuation Indicators}
\label{subsec:pabha}
\begin{figure*}[t]
\epsscale{1.1}
\plotone{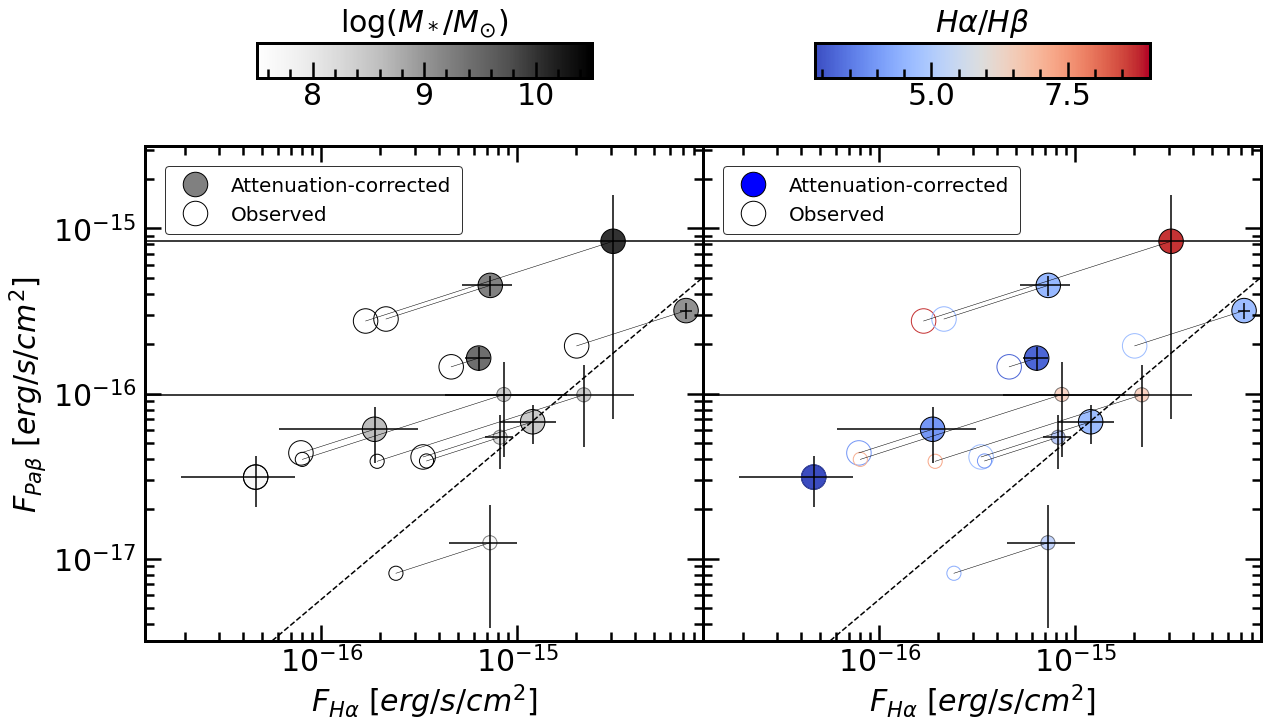}
\caption{\Pab\ and \Ha\ fluxes for the 11 galaxies in our sample with matching TKRS optical spectroscopy, color-coded by stellar mass (\textit{left}) and Balmer decrement (\textit{right}). The dashed gray line indicates $\Pab/\Ha = 1/17.6$, appropriate for Case B recombination with $T=10^4$~K and $n_e=10^4$~cm$^{-3}$ \citep{oste89}. Open circles show uncorrected fluxes and filled circles are dust-corrected fluxes, calculated using the observed Balmer decrement and a \citet{calz00} attenuation curve. Larger symbols represent objects in our primary sample with \Pab\ SNR>3, and smaller symbols represent objects in our secondary sample with reliable redshifts and marginal \Pab\ 1<SNR<3. About half of the objects have attenuation-corrected ratios of $\Pab/\Ha$ which are significantly larger than the expected ratio, over a wide range of stellar mass and Balmer decrement. This picture is supported by a scenario where the small aperture size of the TKRS measurements and the assumptions of invariant equivalent widths leads to potentially under-measured Balmer line fluxes. Some objects have large uncertainties in attenuation-corrected \Ha\ flux due to highly uncertain \Hb\ flux measurements.
\label{fig:pabhaflux}}
\end{figure*}

\begin{figure*}[t]
\epsscale{1.1}
\plotone{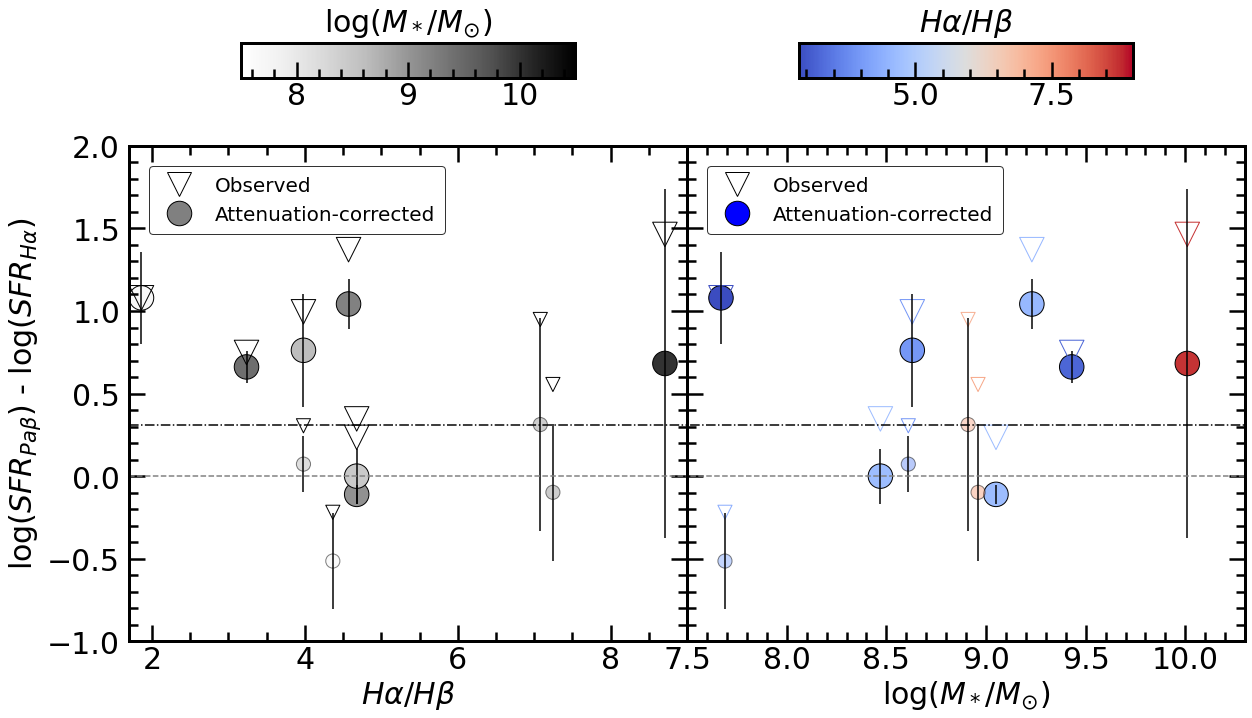}
\caption{The relation of $SFR_{{\rm Pa}\beta}/SFR_{{\rm H}\alpha}$ to Balmer decrement and stellar mass. Open downward-facing triangles show the observed SFRs, while filled symbols show the attenuation-corrected fluxes using the observed Balmer decrement and a \citealt{calz00} attenuation model (assuming an intrinsic Balmer decrement of $\Ha/\Hb=2.86$). Larger symbols represent objects in our primary sample with \Pab\ SNR>3, and smaller symbols represent objects in our secondary sample with reliable redshifts and marginal \Pab\ 1<SNR<3. The \Pab\ dust corrections are generally a factor of two or less, while the \Ha\ dust corrections are often a factor of several or more, which explains the observed ratios exceeding the attenuation-corrected ratios. Even after dust correcting according to the Balmer decrement, many of the galaxies have dust-corrected \Pab\ fluxes that are significantly greater than the expected ratio  of $\Pab/\Ha = 1/17.6$. This indicates that grism-based \Pab\ picks up star formation missed by slit-based optical emission-line SFR indicators. We fit the attenuation-corrected points and observe an excess of \Pab\ star-formation consistent with a constant offset of $\sim0.31$ dex.
\label{fig:pabhasfr}}
\end{figure*} 

We compare the \Pab\ and \Ha\ fluxes and SFRs to investigate if the near-IR \Pab\ emission line reveals star formation that is otherwise hidden in optical emission. Figure \ref{fig:pabhaflux} shows the \Pab\ and \Ha\ fluxes for galaxies in our sample that have optical spectroscopy from TKRS in GOODS-N \citep{wirt04}. Open symbols show the observed fluxes, while filled symbols show the attenuation-corrected fluxes using the Balmer decrement and a \citealt{calz00} attenuation model (assuming an intrinsic Case B Balmer decrement of $\Ha/\Hb=2.86$). The \Pab\ dust corrections are generally a factor of two or less, while the \Ha\ dust corrections are often a factor of several or more. Even after correcting for dust attenuation according to the Balmer decrement, many of the galaxies have dust-corrected \Pab\ fluxes that are significantly greater than the expected ratio for \Ha/\Pab\ (17.6/1, \citealp{oste89}). This suggests that the Balmer decrement is likely to underestimate the dust attenuation affecting \Ha\ in many of our galaxies. In regions of high optical depth to \Hb\ in particular, we have highly uncertain attenuation corrections, which leads to highly uncertain attenuation-corrected \Ha\ fluxes.

Figure \ref{fig:pabhasfr} shows the log ratio of the \Pab\ and \Ha\ SFRs. As in Figure \ref{fig:pabhaflux}, open symbols show the uncorrected SFRs and filled symbols are dust-corrected using the Balmer decrement and a \citet{calz00} attenuation model. Linear regression suggests there are no significant correlations between the $\Pab/\Ha$ ratio and the stellar mass or observed Balmer decrement, with regression slopes consistent with zero with a constant offset of $\sim$0.3 dex. High ratios of the attenuation-corrected $\Pab/\Ha$ are likely to occur if the attenuation is underestimated, i.e. if the Balmer decrement does not measure all of the attenuation. Figure \ref{fig:pabhasfr} suggests that, at least within our small sample, Balmer decrements underestimate the attenuation in galaxies spanning a broad range of stellar mass and Balmer decrement.

There is a potential selection effect that could bias our sample toward dustier galaxies because we have selected galaxies based on their  \Pab\ flux. One potential remedy to this is to study the SFRs of the \Ha\ non-detections (similar to the analysis in Figure \ref{fig:survival}), which would select less dusty galaxies than our \Pab-selected sample where the Balmer decrement is more reliably measured. We propose to augment this work with future studies of these \Pab/\Ha\ ratios with much larger datasets, which will be much better suited to perform the statistical analyses necessary to resolve this degeneracy.

We also draw attention the nuance of the different arguments presented in Figure \ref{fig:decrements} (where we compare the total SFRs derived from different means) and Figure \ref{fig:sfrcomparison} (where we compare the Pa-beta/H-alpha ratios against galaxy dust attenuation estimates). The continuum SFR indicators (e.g., the rest-UV and far-IR) presented in Figure \ref{fig:sfrcomparison} represent a longer timescale (the UV traces the light from B stars with ages of order $\sim$100 Myr, while the far-IR responds to light from longer lived A and even F stars with ages of order $\sim$1 Gyr; see e.g., \citealt{salim20}) than the timescale probed by Hydrogen recombination lines (which are primarily produced from O stars with ages of $\sim$10 Myr). We show in Figure \ref{fig:sfrcomparison} that the attenuation-corrected UV SFRs agree with the UV+IR ladder SFRs, yet in Figure \ref{fig:decrements} we argue that the differences in SFR derived from the near-IR \Pab\ line and the optical \Ha\ arise from attenuation. We argue these observations are not discrepancies but rather indicate variability in the timescale of star-formation as traced by the different SFR indicators.  This leads us to conclude that the variation between the SFRs derived from the Hydrogen recombination lines (e.g., from \Pab) and those from the UV trace variability in the star-formation histories.  We return to this point in Section \ref{subsec:burstiness}. 

Figure \ref{fig:decrements} directly compares the observed $\Pab/\Ha$ and $\Ha/\Hb$ ratios for the 11 galaxies in our CLEAR sample with optical spectra from TKRS \citep{wirt04}. The lines indicate the expected ratios for \citet{calz00}, SMC \citep{gord03}, and Milky Way \citep{fitz99} attenuation models, using intrinsic Case~B ratios of $\Ha/\Hb=2.86$ and $\Pab/\Ha=1/17.6$. Most (8/11) galaxies have line ratios that are broadly (within $<$3$\sigma$) consistent with the expectation from \cite{calz00}, but three have significantly larger $\Pab/\Ha$ ratios than expected from their $\Ha/\Hb$ ratios. As in Figures \ref{fig:pabhaflux} and \ref{fig:pabhasfr}, we interpret these galaxies as having $\Ha/\Hb$ ratios that underestimate the true attenuation. In the galaxy with the highest $\Ha/\Hb$ (upper right of Figure \ref{fig:decrements}), this is somewhat expected since the Balmer decrement is likely to be inaccurate due to gas which is optically thick to \Hb\ emission. This object has a significant dust lane, visible in the images in Figure \ref{fig:images}. The other two galaxies might similarly have Balmer decrements that probe only optically thin gas, with an additional ISM component that is optically thick to \Hb\ (and possibly \Ha) but optically thin to \Pab\ emission.

\subsection{Nebular \Pab/\Ha\ and Continuum Attenuation}
 
\begin{figure*}[t] 
\epsscale{1.1}
\plotone{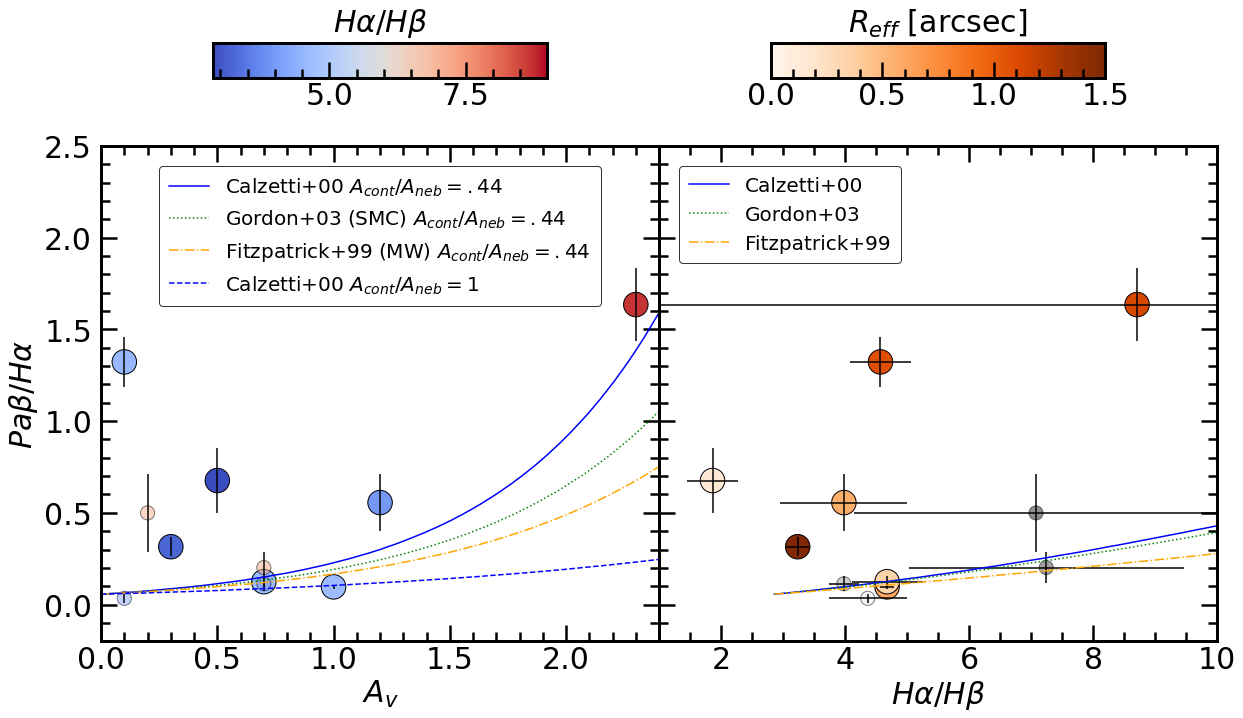}
\caption{\textit{Left:} The relation between $\Pab/\Ha$ ratios and the 3D-HST $A_v$ \citep{momc16} and \cite{barr19} UV attenuation $A_{280}$ for 11 galaxies in our sample with public optical spectroscopy from TKRS \citep{wirt04}. The blue, green, and orange lines show the \cite{calz00}, \cite{gord03}, and \cite{fitz99} attenuation curves with a stellar to nebular attenuation ratio of 0.44, along with another \cite{calz00} attenuation curve with a stellar to nebular attenuation ratio of 1. Larger symbols represent objects in our primary sample with \Pab\ SNR>3, and smaller symbols represent objects in our secondary sample with reliable redshifts and marginal \Pab\ 1<SNR<3. \textit{Right:} The relation between $\Pab/\Ha$ and the Balmer decrement for the same 11 objects with Balmer line fluxes. The blue, green, and orange lines indicate the expected ratios using intrinsic Case B ratios of $\Ha/\Hb=2.86$ and $\Pab/\Ha=1/17.6$, and \citet{calz00} \cite{gord03}, and \cite{fitz99} attenuation models. Eight of the 11 points have line ratios within 3$\sigma$ consistent with the expectation. Our sample includes at least one highly dusty galaxy (in the upper right) for which $\Ha/\Hb$ cannot reliably measure dust attenuation due to high optical depth to Balmer emission.
\label{fig:decrements}}
\end{figure*} 

The left panel of Figure \ref{fig:decrements} compares the nebular attenuation measured by $\Pab/\Ha$ with the continuum $V$-band attenuation $A_V$. We also show \citet{calz00}, Milky Way \citep{fitz99}, and SMC \citep{gord03} attenuation curves for a stellar to nebular attenuation ratio of 0.44, as well as the \citet{calz00} curve for equal stellar and nebular attenuation. All models assume an intrinsic $\Pab/\Ha=1/17.6$. Most (8/11) of the galaxies have observed $\Pab/\Ha$ ratios that are $<3\sigma$ consistent with the dotted line, albeit with significant excess scatter that suggests a large diversity of stellar to nebular attenuation ratios. Three of the galaxies have significantly larger $\Pab/\Ha$ ratios than expected for their $A_V$, and these are the same three galaxies with larger $\Pab/\Ha$ than expected for their Balmer decrements. These galaxies are likely to have measured $A_V$ values which underestimate the attenuation, and/or have enshrouded star-forming regions that are optically thick to optical line and continuum emission but apparent in the near-IR \Pab\ line.

\subsection{Issues with Balmer-line measurements from TKRS slit spectroscopy}
\label{subsec:pabha_issues}
The small size of our sample, with only 11 galaxies that have both rest-optical spectroscopy for \Hb\ and \Ha\ along with rest-IR spectroscopy for \Pab, makes it unclear if our observations are representative of the nebular attenuation properties in the broader population of galaxies. Our \Pab-selected sample is also likely to over-represent high ratios of $\Pab/\Ha$: starting from a \Ha- or \Hb-selected sample would likely result in a different distribution of $\Pab/\Ha$. However, we find evidence that at least some galaxies (spanning the range of stellar mass and observed $\Ha/\Hb$) have Balmer decrements that underestimate the attenuation and miss star-formation that is otherwise revealed by \Pab.

Another potential issue with the direct comparison of \Pab\ emission from CLEAR and Balmer emission from TKRS is the potential for slit losses in the Keck observations. The TKRS disk-integrated Balmer line fluxes are estimated from TKRS spectra in a way that attempts to account for slit losses by assuming that the emission-line equivalent widths are invariant across the stellar disk; see Section 2 of \cite{wein07}.

This assumption likely fails for extended objects, evident from the objects with the largest \Pab/\Ha\ discrepancies from the expected models in the right panel of Figure \ref{fig:decrements} also being some of the most extended objects in the sample (see Figure \ref{fig:images}).  This is important as the slit-width used by TKRS is 1 arcsecond, while the galaxy isophotes extend over many arcseconds.  The assumption of invariant emission-line equivalent widths across extended galaxies with large color gradients can lead to significantly underestimated Balmer emission for galaxies with central dust lanes and/or higher emission-line equivalent widths in their outer regions.

In this section we have discussed the advantages of using comparisons of \Pab\ to emission-line tracers and continuum indicators to show dust attenuation missed by these other methods. We showed that the ratio $\Pab/\Ha$ is a valuable attenuation indicator in moderate to dusty galaxies. A more nuanced analysis of these issues and the benefits of a three emission-line attenuation model using the CLEAR sample is discussed in a companion work \cite{Prescott2022}.

\section{\Pab\ as a Star-Formation Rate Indicator}\label{sec:sfr}

Following the \citet{kenn12} SFR relation for \Ha\ and $\Ha/\Pab=17.6$ (assuming Case B recombination, $T=10^4$~K, and $n_e=10^4$~cm$^{-3}$; \citealp{oste89}), SFR is calculated from \Pab\ as:
\begin{equation}\label{eq:sfrpab}
  \log({\rm SFR}_{{\rm Pa}\beta}) [M_\odot/\mathrm{yr}] = \log[L(\Pab)] - 40.02 +0.4A_{{\rm Pa}\beta}
\end{equation}

This equation includes an attenuation correction for the \Pab\ emission, $A_{{\rm Pa}\beta}$, calculated from the measured continuum attenuation $A_V$ with a \cite{calz00} attenuation curve and a stellar-to-nebular attenuation ratio of 0.44 \citep{calz94}. In the following subsections, we compare \Pab\ SFRs measured from these equations with continuum and \Ha\ SFR estimates.
\subsection{\Pab\ and Continuum SFR Indicators}
\begin{figure}[t]
\epsscale{1.15}
\plotone{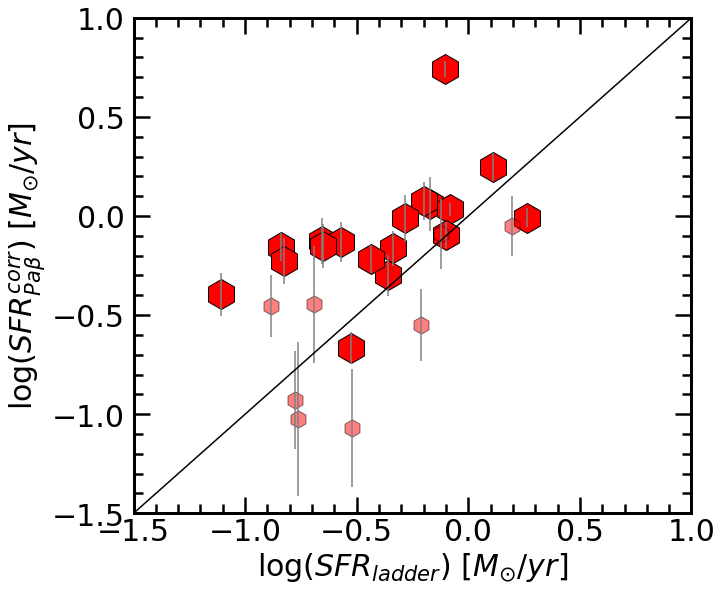}
\caption{The relation between attenuation-corrected \Pab\ SFR and UV+IR `ladder' star-formation rates for galaxies in our sample. Larger symbols represent objects in our primary sample with \Pab\ SNR>3, and smaller symbols represent objects in our secondary sample with reliable redshifts and marginal \Pab\ 1<SNR<3. The solid black line indicates the one-to-one relation. We note that the attenuation-corrected \Pab\ SFRs are  at lease $3 ~\sigma$ greater than the ladder SFRs for 18 of the 29 objects in our sample. The object with the highest attenuation-corrected \Pab\ SFR is GN 34456, which has a significant dust lane (see Figure \ref{fig:images}).
\label{fig:pabladder}}
\end{figure}
\begin{figure*}[t]
\epsscale{1.15}
\plotone{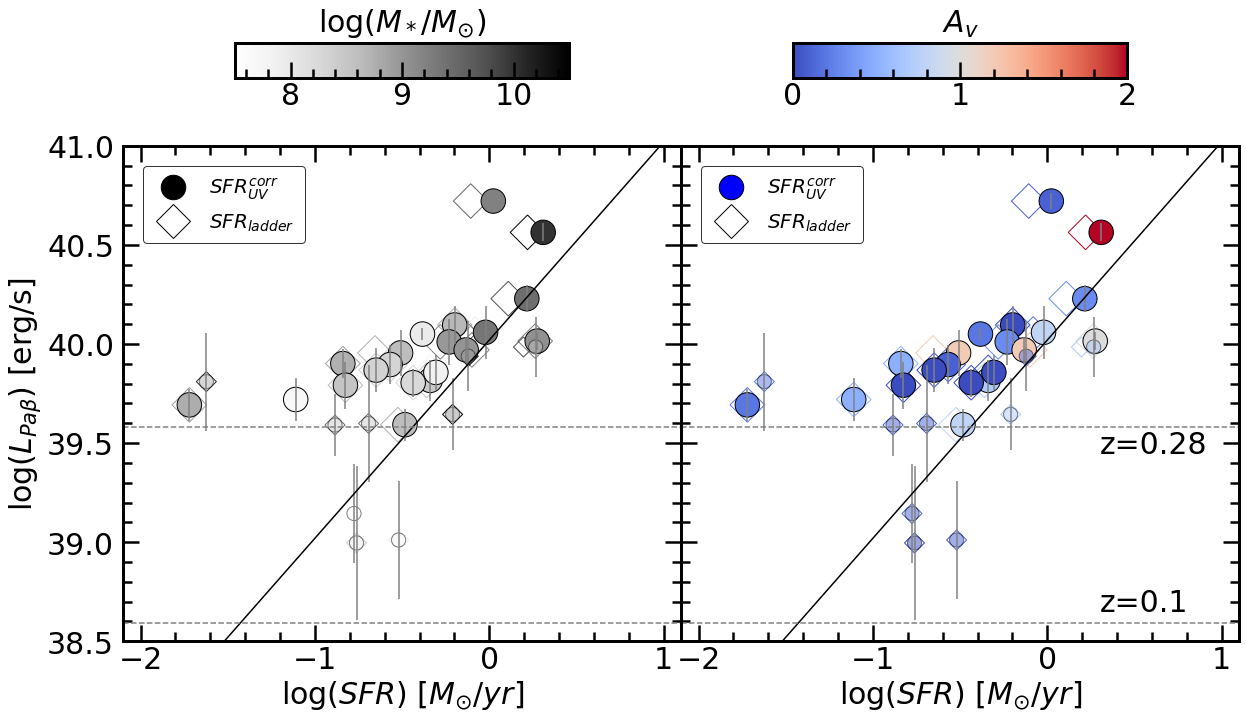}
\caption{The relation between \Pab\ luminosity and continuum star-formation rates for galaxies in our sample, color coded by mass (\textit{left}), and continuum $A_v$ (\textit{right}). Filled circles correspond to the attenuation-corrected UV SFRs and open diamonds correspond to UV + IR ladder SFRs for the twelve galaxies with well-detected IR emission \citep{barr19}. Larger symbols represent objects in our primary sample with \Pab\ SNR>3, and smaller symbols represent objects in our secondary sample with reliable redshifts and marginal \Pab\ 1<SNR<3. The solid black line indicates the relation between \Pab\ luminosity and SFR calculated using Equation \ref{eq:sfrpab} \citep{kenn12}. The dashed gray lines represent the 1-sigma detection limits at redshifts $z=0.28$ and $z = 0.1$. The SFR measured from the \Pab\ luminosity is higher than the attenuation-corrected UV SFR in galaxies with higher attenuation, and there is more apparent scatter between the two SFRs in galaxies with low stellar mass.
\label{fig:pabuv}}
\end{figure*}

\begin{figure*}[t]
\epsscale{1.15}
\plotone{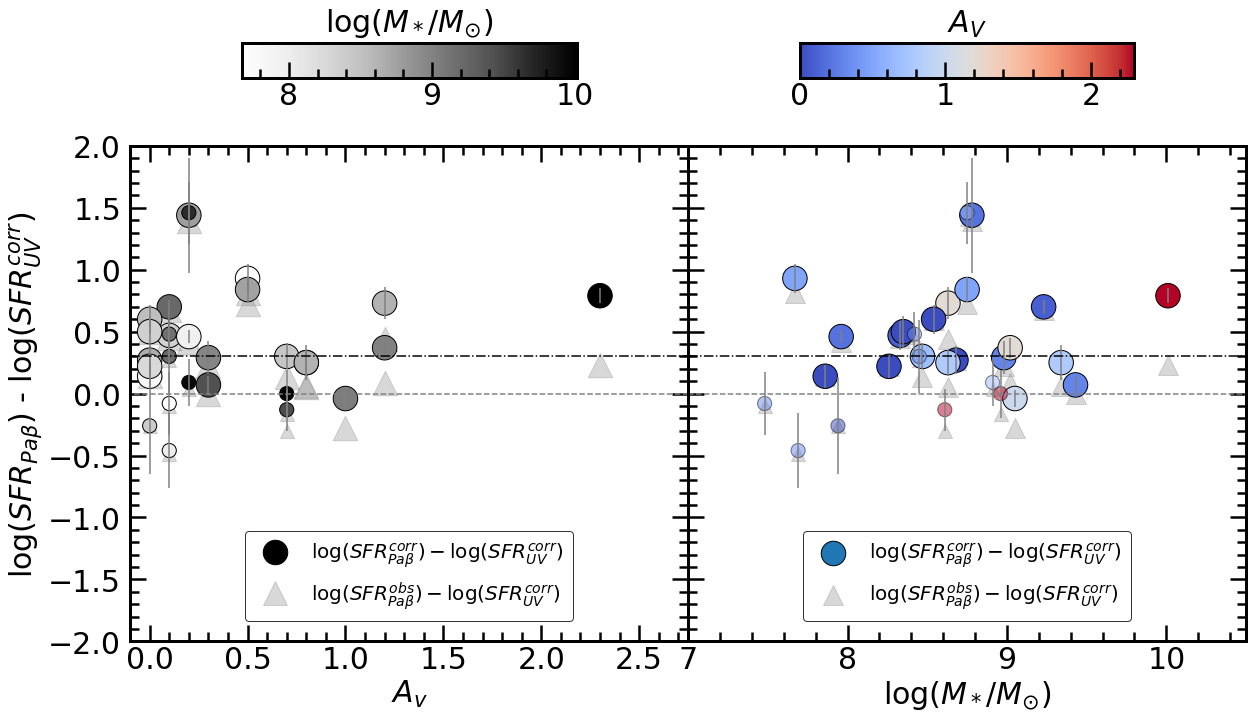}
\caption{The log ratio of the \Pab\ and attenuation-corrected UV SFRs with stellar mass (\textit{left}) and continuum $A_v$ (\textit{right}). Solid circles represent attenuation-corrected \Pab/UV SFR ratios, and hollow circles correspond to the same objects without attenuation-corrected \Pab. Larger symbols represent objects in our primary sample with \Pab\ SNR>3, and smaller symbols represent objects in our secondary sample with reliable redshifts and marginal \Pab\ 1<SNR<3. We perform a linear regression fit to the attenuation-corrected \Pab/UV SFR ratios in each panel, finding no significant correlations between the \Pab/UV SFR ratios and stellar mass or $A_V$. The black dash-dot line shows the median attenuation-corrected \Pab/UV ratio of $0.3$.
\label{fig:pabuvoff}}
\end{figure*} 

Here we compare \Pab\ SFRs with attenuation-corrected UV continuum SFRs and UV + IR `ladder' SFRs to investigate hidden SFR and star-formation histories (SFHs). The near-IR \Pab\ line is much less affected by dust attenuation than the UV continuum, and so \Pab\ can reveal star-forming regions that are otherwise obscured by dust in light of shorter wavelengths \citep{kenn12}. In addition, \Pab\ (and other hydrogen emission lines) probes recent (<10 Myr) star-formation, while the UV continuum probes star-formation over longer (100-500~Myr) timescales \citep{kenn12,redd12}. The comparison of \Pab\ and UV continuum SFRs yields an indicator of SFH stochasticity (``burstiness'', e.g., \citealt{guo12, brou19}).

Figure \ref{fig:pabladder} shows the relation between the attenuation-corrected \Pab\ SFRs and the UV+IR ladder SFRs from \cite{barr19}. We note that for 18 of the 29 objects in our sample, the \Pab\ SFR is $>$3$\sigma$ higher than the UV+IR ladder SFRs. The object with the highest attenuation-corrected \Pab\ SFR is GN 34456, which has a significant dust lane (see Figure \ref{fig:images}).

Figure \ref{fig:pabuv} shows the relation between \Pab\ luminosity and the attenuation-corrected UV SFR from the CANDELS/SHARDS multiwavelength catalog \citep{barr19}, with three panels color-coded by stellar mass and $V$-band attenuation. Twelve  of the galaxies in our sample have 24~$\mu$m detections, and their UV + IR ladder SFRs are shown as open diamonds \citep{wuyt11}. Both SFRs correlate with stellar mass, as expected given the well-known star-formation mass sequence of galaxies \citep{noeske07,whit12}, but there is more apparent scatter between the two SFR indicators in low-mass galaxies. The \Pab\ luminosity also tends to be greater than expected from the UV SFR in galaxies with steep UV slopes and high attenuation.

Figure \ref{fig:pabuvoff} shows the ``excess'' \Pab\ SFR compared to the attenuation-corrected UV SFR, quantified as $\log\frac{SFR_{Pa\beta}}{SFR_{UV}^{corr}}$, with stellar mass and $V$-band attenuation $A_V$. Our \Pab-selected sample is generally only sensitive to galaxies with \Pab\ SFR similar to or greater than the UV SFR. We fit the attenuation-corrected detections (colored points) in each panel using linear regression, as implemented by the \texttt{linmix} \citep{kell07} Python package. We find no significant correlations (with a slope $>$3$\sigma$ different from zero) between the $\log\frac{SFR_{Pa\beta}}{SFR_{UV}^{corr}}$ ratio and mass or $A_V$. The \Pab\ SFR ``excess'' is instead consistent with a constant offset of $\sim0.3$ dex over the attenuation-corrected UV SFR.

Some galaxies in our sample tend to have \Pab\ SFRs that are $\sim$1-2 orders of magnitude greater than the attenuation-corrected UV SFRs (see Figure \ref{fig:pabuvoff}). These galaxies may have dust-enshrouded star formation that is not seen in UV light (for example, behind high optical depths) and is not accounted for by the attenuation correction of the UV SFR. \Pab\ emission can escape these dusty star-forming regions that are optically thick to UV light, revealing star-formation that is hidden at shorter wavelengths.

\begin{figure*}[h]
\epsscale{1.15}
\plotone{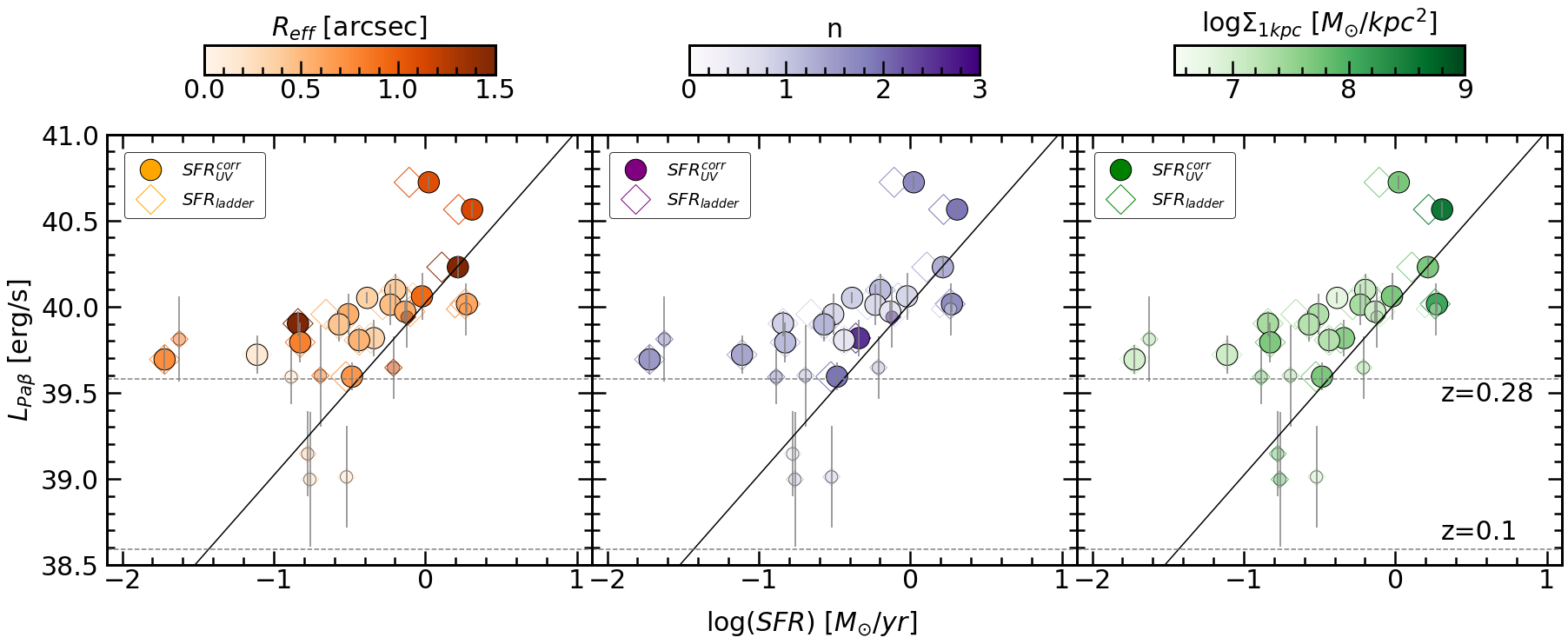}
\caption{The relation between \Pab\ luminosity and attenuation-corrected UV star-formation rates for galaxies in our sample, color coded by effective radius (\textit{left}), S\'ersic index (\textit{center}) and $\Sigma_{1kpc}$ (\textit{right}). Filled circles correspond to the attenuation-corrected UV SFRs and open diamonds correspond to UV + IR ladder SFRs for the twelve galaxies with well-detected IR emission \citep{barr19}. Larger symbols represent objects in our primary sample with \Pab\ SNR>3, and smaller symbols represent objects in our secondary sample with reliable redshifts and marginal \Pab\ 1<SNR<3. As in Figure \ref{fig:pabuv}, the solid black line is the relationship between \Pab\ luminosity and SFR following Equation \ref{eq:sfrpab} \citep{kenn12}, and the dashed gray lines represent the 1-sigma detection limits at redshifts $z=0.28$ and $z = 0.1$. None of these quantities have an apparent correlation with the ratio of \Pab\ to attenuation-corrected UV star-formation rates, shown quantitatively in Figure \ref{fig:pabuvoff_morph}.
\label{fig:pabuv_morph}}
\end{figure*} 

\begin{figure*}[h]
\epsscale{1.15}
\plotone{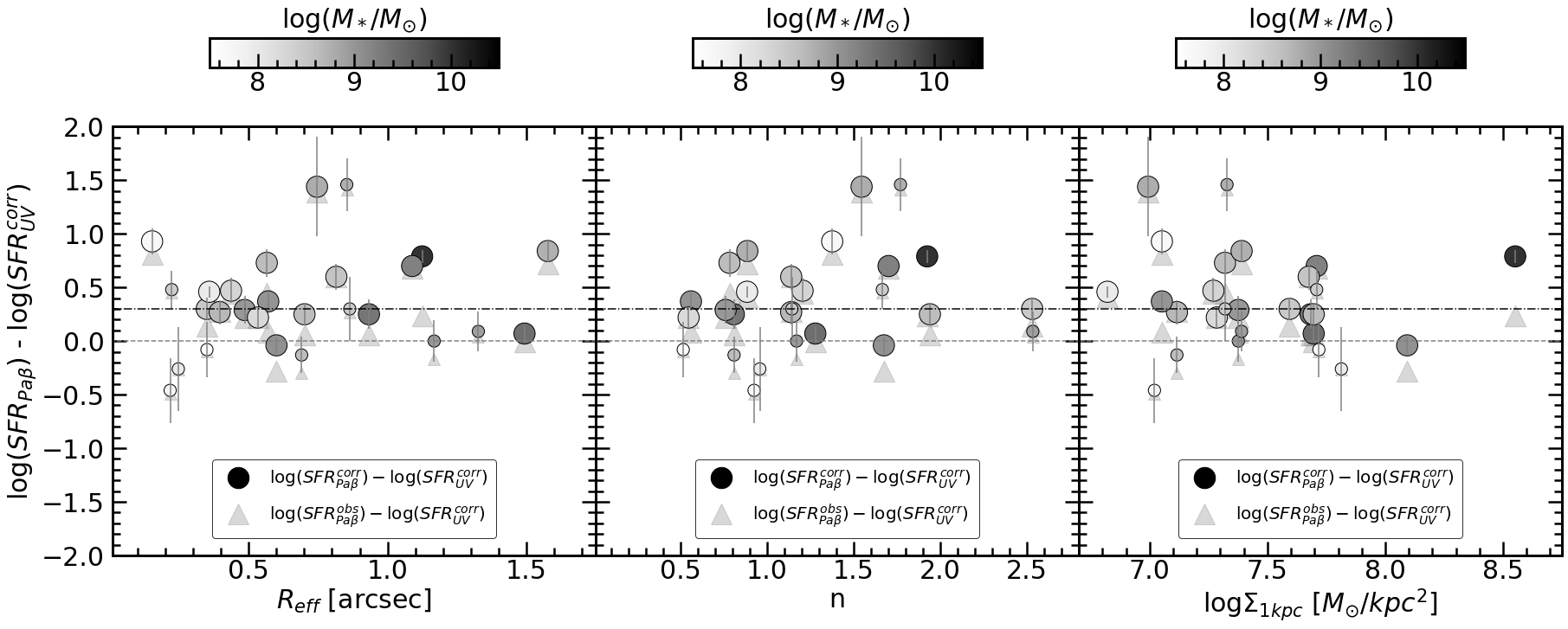}
\caption{The log ratio of the \Pab\ and attenuation-corrected UV SFRs with galaxy effective radius (\textit{left}) S\'ersic index (\textit{center}), and central density $\log\Sigma_{1\text{kpc}}$ (\textit{right}). Solid circles represent attenuation-corrected \Pab/UV ratios, and open circles represent the same objects without attenuation-corrected \Pab. Larger symbols represent objects in our primary sample with \Pab\ SNR>3, and smaller symbols represent objects in our secondary sample with reliable redshifts and marginal \Pab\ 1<SNR<3. We find no significant correlations between the attenuation-corrected \Pab/UV ratio and effective radius, S\'ersic index, or central density. The black dash-dot line shows the median attenuation-corrected \Pab/UV ratio of $0.3$.
\label{fig:pabuvoff_morph}}
\end{figure*}

There is a broad range of \Pab\ SFR excess in Figure \ref{fig:pabuvoff} for galaxies of varying stellar masses and $A_V$. However, our measured \Pab/UV SFRs may be influenced by the \Pab\ detection limits, which are mostly closer to a ratio of \Pab\ to UV SFRs of unity at low stellar mass. Our \Pab-selected study is generally only sensitive to bursty star-formation that occurred within the last 10~Myr, detectable as \Pab\ emission (with ${\rm SFR}_{\Pab}>{\rm SFR_{UV}}$). In contrast, a burst of star-formation occurring 10-100~Myr ago would lead to ${\rm SFR}_{\Pab}<{\rm SFR_{UV}}$ that is generally not detectable in our sample (as shown by the black upward facing triangles that indicate the \Pab\ flux limit for each galaxy in Figure \ref{fig:pabuvoff}). Thus we interpret the scatter of \Pab/UV SFRs in Figure \ref{fig:pabuvoff} as consistent with bursty star-formation in some galaxies. Furthermore,the \Pab\ detection limits are consistent with an undetected population of low-mass galaxies with ${\rm SFR}_{\Pab}<{\rm SFR_{UV}}$ that represent bursty star-formation occurring $>$10~Myr ago, analogous to detected ${\rm SFR}_{\Pab}>{\rm SFR_{UV}}$ galaxies representing bursts within the last 10~Myr. We discuss this further in Section \ref{subsec:burstiness}.

Figure \ref{fig:pabuv_morph} shows the \Pab\ luminosity and attenuation-corrected UV SFR color-coded by galaxy size, S\'ersic index (measured by \citealt{van12}), and central density $\Sigma_{1kpc}$ calculated from these values and their stellar masses. Figure \ref{fig:pabuvoff_morph} shows the log ratio of the \Pab\ to UV SFR versus the same morphology quantities. There are no significant correlations between the ratio of the two SFRs with galaxy size, S\'ersic index, or $\Sigma_{1kpc}$. Instead, we observe the same constant vertical offset of $\sim0.3$ dex. We conclude that galaxy morphology does not play a dominant role in the ratio between the \Pab\ and UV SFR of a galaxy. We discuss this further in the next subsection.

We note the average (median) attenuation-corrected \Pab\  ``excess'' ($\log\frac{SFR_{Pa\beta}}{SFR_{UV}^{corr}}$) ratio for our sample of 29 galaxies is 0.3. We show this by the horizontal black dash-dot line in Figures \ref{fig:pabuvoff} and \ref{fig:pabuvoff_morph}. We have also performed a stack of 44 objects in CLEAR with $0.2<z<0.3$ (both detections and non-detections) and found a stacked \Pab\ luminosity of $2.5\pm 2.0 \times 10^{39}$ erg/s, which corresponds to a $SFR_{\Pab} \sim 0.21 \pm 0.16~M_\odot$/yr. For this sample of 44 objects, the median attenuation-corrected UV continuum SFR is $0.44~M_\odot$/yr, which is comparable to that of \Pab\ to within about 1$\sigma$. This stacking analysis is consistent with the conclusions of the following subsection: namely that the Pab and attenuation-corrected UV SFRs are broadly consistent (especially when considering non-detections), but with large scatter in the ratio of SFRs among individual galaxies.

Our sample of \Pab\ detections might be biased toward high \Pab/UV ratios, since fainter \Pab\ emission would be undetected and excluded from the sample. In the following section we perform a survival analysis of the \Pab/UV ratio with a sample of 152 \Pab\ non-detections and our 29 \Pab\ detections in the CLEAR survey to remedy these sample-selection biases.

Future studies with deeper grism detections of Paschen-lines (see discussion of \textit{JWST} surveys in the summery) will also add valuable information on analyses of this kind.

\subsection{Burstier Star-Formation at Low Stellar Mass}\label{subsec:burstiness}

The different star-formation timescales probed by hydrogen recombination lines ($\sim$5~Myr) and near-UV continuum emission ($\sim$100~Myr) means that their comparison can be used to indicate the burstiness of star-formation. Observations have long shown that the average ratio of Balmer-line to UV SFRs decreases at lower stellar mass \citep{sull00,bose09,lee09,guo16}. The observations are best explained by an increasing importance of bursty star-formation occurring on timescales of tens of Myr, in which star-formation that occurred 10-100~Myr ago is detected in UV emission but not in hydrogen emission lines \citep{weis12,hopk14,shen14,sparr17}. But \citet{brou19} notes that comparisons between Balmer-line and UV SFRs can be biased by uncertainties in the amount of dust attenuation (especially the UV shape of the attenuation law and the ratio of nebular to continuum attenuation).

\begin{figure*}[h] 
\epsscale{1.1}
\plotone{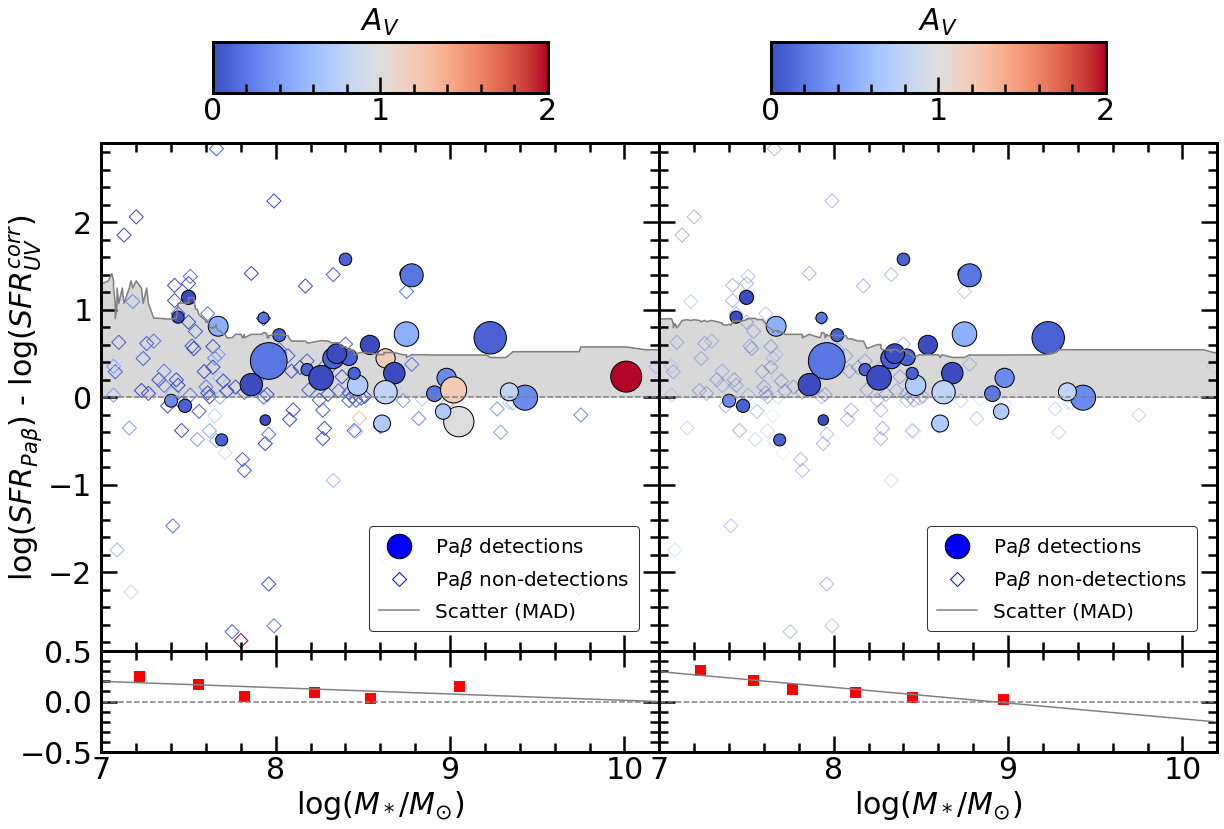}
\caption{Survival analysis of the log ratio of \Pab\ and attenuation-corrected UV SFRs to stellar mass (similar to the left panel of Figure \ref{fig:pabuvoff}). The solid circles are all of the objects in our sample \Pab\ detections of SNR>1. Open diamonds are randomly generated data from these non-detections using a half-normal distribution with standard deviation equal to the 1$\sigma$ \Pab\ upper limits of the non-detections. Both detections and the randomly generated survival analysis points are color-coded by $A_{V}$.  The gray lines show rolling median absolute deviations of the sample and survival analysis points. The right panel shows the same analysis for the subsample of galaxies with $A_{V} < 1$. The survival analysis particularly populates the lower-left of the plot where we cannot detect \Pab-emitting galaxies. In order to quantify the scatter, we resample each point 200 times and take the standard deviation for in each of six stellar mass bins. We then fit the standard deviations for each resampling and find a median slope of $m = -0.063 \pm 0.039$ (left panel) and  $m = -0.151 \pm 0.048$ (right panel), indicating higher scatter at lower stellar mass with $\approx 1.6\sigma$ and $\approx3.1\sigma$ significance, respectively. The higher significance in the right panel, where moderate and highly dusty objects are eliminated, indicates that the scatter in the \Pab/UV ratio is not due to variable attenuation. We interpret this higher scatter at lower stellar mass to be evidence for burstier SFHs at lower stellar mass, consistent with previous work \citep{brou19,guo16,weis14}.
\label{fig:survival}}
\end{figure*} 

Because the 29 \Pab-detected galaxies shown in Figure \ref{fig:pabuvoff} are selected by their \Pab\ fluxes, the sample might be biased to high \Pab/UV ratios. To test this bias, we identify the sample of \Pab\ non-detections in CLEAR in the same redshift range and perform a survival analysis of the \Pab/UV ratio with this total sample of detections plus non-detections. We perform a survival analysis instead of a stacking analysis to preserve the scatter and eliminate loss of information from binning \citep{feig12}.

To test the possibility that low-mass galaxies are just as likely to have suppressed \Pab/UV as enhanced \Pab/UV, we perform a survival analysis of the relation of the \Pab\ to attenuation-corrected UV SFR ratio to stellar mass in Figure \ref{fig:survival}. We include all objects with \Pab\ SNR>1, and generate survival analysis expectations for the 152 \Pab\ non-detections in CLEAR. The inclusion of the non-detections will populate the low \Pab/UV regime, where our \Pab-selected sample is biased against. The survival analysis reveals the large undetected population of low-\Pab/UV galaxies with a distribution (especially for galaxies with less attenuation) that is broadly consistent with previous work (cf. Figure 4 of \citealt{weis12}).

For each non-detection we randomly generate expectation values from a half-normal distribution with standard deviation equal to the 1$\sigma$ upper bounds of each of the non-detections (open diamonds). The survival analysis does not have a significant dependence on the assumed prior distribution of the resampled points. The gray lines show rolling median absolute deviations of the sample and survival analysis points.  We perform the following analyses both for the full sample of detections and randomly sampled non-detections (left panel of Figure \ref{fig:survival}) and again for detections and non-detections for relatively unattenuated galaxies with $A_{V} < 1$ (right panel of Figure \ref{fig:survival}). 

We fit the scatter of the \Pab\ to UV SFR ratio in 6 bins of stellar mass for the \Pab\ detections and the survival analysis draws of the non-detections. We resample the measurements and the survival analysis draws 200 times and calculate the standard deviation of the distribution of SFR ratios measured from the resampled points in each bin. We then fit the standard deviations of SFR ratios across the six bins and find a best-fit slope of $m = -0.063 \pm 0.039$ (left panel) and  $m = -0.151 \pm 0.048$ (right panel), indicating higher scatter at lower stellar mass with $\approx 1.6\sigma$ and $\approx3.1\sigma$ significance, respectively.

The higher significance of the negative correlation between the scatter of the \Pab/UV ratio and stellar mass in the right panel of Figure \ref{fig:survival} indicates that the scatter cannot be due solely to differences in attenuation among these galaxies. Instead, we interpret this as evidence of stochasticity of the star-formation histories. This result of burstier SFHs at lower stellar mass is consistent with previous works using Balmer/UV ratios \citep{weis12,hopk14,shen14,sparr17}.

The \Pab\ line should offer a clearer approach to measuring the burstiness of star-formation. Because \Pab\ is much less affected by attenuation than the Balmer lines, we will be able to measure star-formation histories without dealing with the potential biases due to dust, a significant problem for optical emission-line to UV ratio star-formation history studies in the literature \citep{brou19}.

\section{Summary and Conclusions}\label{sec:conclusions}

We have analyzed a sample of 29 low-redshift ($z<0.287$) \Pab\ emitting galaxies. The galaxies of our sample have been divided into two samples: a primary sample selected such that the $\hst$ G141 grism detects \Pab\ with an observed frame wavelength of $\lambda \leq 16500\text{\AA}$ with a signal-to-noise ratio of $\sigma \geq 3$ (20 galaxies), and a secondary sample of 9 galaxies with reliable spectroscopic redshifts from other lines (from either ground-based optical spectroscopy or from \Ha\ emission in the G102 spectrum) that have \Pab\ detected with ${\rm SNR}>1$. We also required that the objects in our sample have minimally contaminated 2-D spectra by visual inspection.

We show that \Pab\ is a valuable indicator of star-formation rates and star-formation histories when compared to more widely used SFR indicators such as \Ha\ flux and continuum emission, especially in moderately-dusty to highly-dusty galaxies. \Pab\ as an indicator of SFR serves as a solution to the issues with dust attenuation experienced by SFRs based on optical emission line tracers such as $\Ha$.
 
Our study of these \Pab\ emitting galaxies provides two primary findings:
\begin{itemize}
    \item SFRs calculated from \Pab\ probe a shorter timescale than probed by continuum emission, so we can draw conclusions about the star-formation histories of galaxies by comparing the two. We consider that the scatter in the \Pab/UV SFR ratio is due to the variability of  dust attenuation, and rule this out based on observations of the full optical--to--mid-IR spectral energy distributions.  Rather, we argue that the scatter in the \Pab/UV SFR ratio indicates increased stochasticity of the star-formation histories, which increases with decreasing stellar mass. This is substantiated by our survival analysis analysis, and agrees with other, independent observations in previous work.
    
    \item $\Pab/\Ha$ ratios serve as a valuable indicator of dust attenuation when we compare these to the Balmer decrement and continuum attenuation estimates, notably in moderately to severely dusty galaxies. $\Pab/\Ha$ has the same insensitivity to nuisance parameters such as metallicity, temperature, and density as the Balmer decrement, but does not risk miscalculating attenuation for ISM regions optically thick to \Hb. The Balmer decrement also develops large uncertainties in only moderately dusty galaxies due to poorly constrained \Hb\ emission.
\end{itemize}

Our results motivate future IR observations of Paschen series lines for measuring star-formation rate. The \textit{James Webb Space Telescope} (\textit{JWST}) will reach a flux limit that is an order of magnitude fainter than our CLEAR data for similar exposure times, enabling detection of fainter Paschen-line emission in low stellar mass galaxies where our work has to rely on survival analysis. In addition, the broad 1-5~$\mu$m spectroscopic coverage of \textit{JWST} includes the \Paa\ line, which is twice as bright than \Pab, for galaxies over $z<1.65$. Future \textit{JWST} observations of Paschen-line emission in galaxies are likely to reveal a much more complete picture of star-formation and bursty formation histories, especially in galaxies with significant dust attenuation.

NIRCam grism slitless spectroscopy of the Balmer lines in conjunction with \Pab\ or \Paa\ can offer more accurate Paschen-to-Balmer dust attenuation measurements, without the need for relying on ground-based optical measurements. This would offer a more accurate census of dust attenuation in highly dusty galaxies without making assumptions about slit losses which may fail for certain morphologies.

\software{linmix \citep{kell07}, \texttt{grizli} pipeline \citep{bram08}, FAST \citep{krie09}, EAZY \citep[]{bram08, wuyt11}, GALFIT \citep{peng10}, Astropy \citep{astr13},
Matplotlib \citep{hunt07}}

\acknowledgements 

The authors wish to thank our colleagues in the CLEAR collaboration for their work on this project, and their assistance and support.  The authors thank the anonymous referee for valuable feedback and constructive comments that improved the quality and clarity of the analysis and interpretation in this paper. We would like to give special thanks to Rob Kennicutt, Kristian Finlator, and Moire Prescott for insightful discussion in the final stages of this paper.

This work is based on data obtained from the Hubble Space Telescope through program number GO-14227. Support for Program number GO-14227 was provided by NASA through a grant from the Space Telescope Science Institute, which is operated by the Association of Universities for Research in Astronomy, Incorporated, under NASA contract NAS5-26555. NJC, JRT, and BEB acknowledge support from NSF grant CAREER-1945546 and NASA grants JWST-ERS-01345 and 18-2ADAP18-0177. NJC also acknowledges support from NASA/\textit{HST} AR 16609. VEC acknowledges support from the NASA Headquarters under the Future Investigators in NASA Earth and Space Science and Technology (FINESST) award 19-ASTRO19-0122. RCS appreciates support from a Giacconi Fellowship at the Space Telescope Science Institute.

\begin{longrotatetable}
\begin{deluxetable*}{llrrrrrrrrrrrrr}
\tablecaption{Sample characteristics and derived quantities \label{tab:sample}}
\tablewidth{700pt}
\tabletypesize{\scriptsize}
\tablehead{
\colhead{Field} & \colhead{ID} & \colhead{RA} & \colhead{Dec} & \colhead{Redshift\tablenotemark{i}} & \colhead{Stellar Mass\tablenotemark{ii}} & \colhead{\Pab\ Flux} & \colhead{$SFR_{UV}^{corr}$\tablenotemark{iii}} & \colhead{$SFR_{ladder}$\tablenotemark{iii}} & \colhead{$\beta$\tablenotemark{iii}} & \colhead{$A_v$\tablenotemark{ii}} & \colhead{$R_{\rm eff}$\tablenotemark{iv}} & \colhead{S\'ersic index\tablenotemark{iv}} & \colhead{$\Sigma_{\rm 1kpc}$} & \colhead{F435W - F775W\tablenotemark{iii}} \\ 
\colhead{} & \colhead{} & \colhead{Deg} & \colhead{Deg} & \colhead{} & \colhead{$\log M_{\odot}$} & \colhead{$10^{-17}$ erg s$^{-1}$ cm$^{-2}$} & \colhead{$M_{\odot}/yr$} & \colhead{$M_{\odot}/yr$} & \colhead{} & \colhead{mag} & \colhead{\arcsec} & \colhead{} & \colhead{$M_{\odot}$ $kpc^{-2}$} & \colhead{mag}
} 
\startdata
      GN1 &  37683 &  189.30609 &  62.36035 &  0.2755 &      8.61 &   3.9 $\pm$ 1.4 &        1.8 $\pm$ 0.3 & 1.57 &     -1.01 &    0.7 &  0.69 $\pm$ 0.02 &      0.81 $\pm$  0.07 &  7.12 & 0.97\\
      GN2 &  19221 &  189.20126 &  62.24070 &  0.1389 &      9.05 &  19.4 $\pm$ 2.1 &        1.9 $\pm$ 0.2 & 1.83 &     -1.18 &    1.0 &  0.600 $\pm$ 0.009 &    1.67 $\pm$  0.05 &  8.09 & 0.91\\
      GN2 &  15610 &  189.21272 &  62.22242 &  0.2008 &      9.43 &  14.5 $\pm$ 2.3 &        1.64 $\pm$ 0.07 & 1.28 &     -1.23 &    0.3 &  1.49 $\pm$ 0.07 &    1.3 $\pm$  0.1 &  7.70 & 1.15\\
      GN2 &  18157 &  189.18229 &  62.23246 &  0.2013 &        8.96 &   3.9 $\pm$ 1.6 &        0.61 $\pm$ 0.095 & 0.62 &     -1.23 &    0.7 &  1.2 $\pm$ 0.2 &     1.2 $\pm$  0.4 &  7.38 & 1.26\\
      GN2 &  21693 &  189.23252 &  62.24847 &  0.28 $\pm$ 0.02 &        8.75 &   2.5 $\pm$ 1.4 &        0.024 $\pm$ 0.002 & 0.02 &     -1.71 &    0.2 &  0.9 $\pm$ 0.3 &    2 $\pm$  1 &  7.33 & 2.03\\
      GN3 &  34456 &  189.33981 &  62.32429 &  0.2113 &        10.01 &  27.5 $\pm$ 2.9 &       2.0 $\pm$ 0.2 & 1.66 &      0.22 &    2.3 &  1.1 $\pm$ 0.2 &        1.9 $\pm$  0.5 &  8.55 & 2.20\\
      GN3 &  34157 &  189.20683 &  62.32120 &  0.2755 &        9.23 &  28.3 $\pm$ 2.7 &        1.05 $\pm$ 0.09 & 0.78 &     -1.70 &    0.1 &  1.1 $\pm$ 0.1 &      1.7 $\pm$  0.4 &  7.71 & 1.21\\
      GN3 &  33397 &  189.17547 &  62.31435 &  0.25 $\pm$ 0.01 &        9.34 &   6.2 $\pm$ 1.9 &        0.95 $\pm$ 0.08 & 0.68 &     -0.46 &    0.8 &  0.9 $\pm$ 0.07 &     0.8 $\pm$  0.1 &  7.68 & 1.66\\
      GN3 &  33511 &  189.23455 &  62.31477 &  0.2535 &      8.63 &   4.4 $\pm$ 1.2 &        0.31 $\pm$ 0.04 & 0.22 &     -1.08 &    1.2 &  0.6 $\pm$ 0.1 &      0.8 $\pm$  0.6 &  7.32 & 1.53\\
      GN3 &  34368 &  189.33853 &  62.32097 &  0.2311 &      8.47 &   4.1 $\pm$ 1.0 &        0.46 $\pm$ 0.01 & 0.44 &     -1.94 &    0.7 &  0.3 $\pm$ 0.2 &      2.5 $\pm$  3.2 &  7.59 & 0.92\\
      GN3 &  34077 &  189.21093 &  62.31770 &  0.25 $\pm$ 0.01 &        8.33 &   4.1 $\pm$ 0.9 &    0.27 $\pm$ 0.04 & 0.27 &     -0.53 &    0.1 &  0.44 $\pm$ 0.03 &    1.2 $\pm$  0.2 &  7.27 & 1.20\\
      GN3 &  35455 &  189.33207 &  62.32867 &  0.2468 &         7.69 &   0.8 $\pm$ 0.6 &    0.30 $\pm$ 0.04 & 0.30 &     -0.97 &    0.1 &  0.22 $\pm$ 0.03 &    0.9 $\pm$  0.5 &  7.02 & 0.64\\
      GN4 &  24611 &  189.35906 &  62.26414 &  0.2662 &        8.91 &   4.0 $\pm$ 1.6 &        0.76 $\pm$ 0.10 & 0.76 &     -1.42 &    0.2 &  1.3 $\pm$ 0.1 &      2.5 $\pm$  0.4 &  7.39 & 1.16\\
      GN5 &  33249 &  189.20772 &  62.31110 &  0.2305 &          7.67 &   3.1 $\pm$ 0.8 &    0.078 $\pm$ 0.007 & 0.08 &     -1.95 &    0.5 &  0.2 $\pm$ 0.03 &   1.4 $\pm$  0.7 &  7.05 & 0.83\\
      GS2 &  45518 &   53.15409 & -27.69793 &  0.282 $\pm$ 0.005 &      8.68 &   4.9 $\pm$ 1.1 &        0.63 $\pm$ 0.09 & 0.63 &     -1.08 &    0.0 &  0.395 $\pm$ 0.004 &  1.14 $\pm$  0.03 &  7.12 & 0.98\\
      GS3 &  37720 &   53.13911 & -27.73031 &  0.1031 &      8.63 &  13.4 $\pm$ 2.5 &        0.33 $\pm$ 0.05 & 0.30 &     -0.85 &    0.8 &  0.701 $\pm$ 0.002 &  1.94 $\pm$  0.01 &  7.70 & 1.03\\
      GS3 &  41882 &   53.17335 & -27.71496 &  0.2501 &      8.98 &   5.9 $\pm$ 1.6 &        0.59 $\pm$ 0.08 & 0.52 &     -0.34 &    0.3 &  0.486 $\pm$ 0.002 &  0.76 $\pm$  0.01 &  7.38 & 1.34\\
      GS3 &  42593 &   53.15580 & -27.71195 &  0.24 $\pm$ 0.02 &        8.42 &   2.1 $\pm$ 0.8 &    0.13 $\pm$ 0.02 & 0.13 &     -1.11 &    0.1 &  0.223 $\pm$ 0.003 &  1.66 $\pm$  0.04 &  7.71 & 1.42\\
      GS3 &  35433 &   53.14153 & -27.74583 &  0.26 $\pm$ 0.02 &        8.45 &   1.9 $\pm$ 1.3 &        0.20 $\pm$ 0.02 & 0.22 &     -1.95 &    0.1 &  0.86 $\pm$ 0.02 &    1.14 $\pm$  0.04 &  7.32 & 1.12\\
      GS4 &  27438 &   53.19393 & -27.78580 &  0.1280 &        8.75 &  23.3 $\pm$ 3.9 &        0.14 $\pm$ 0.02 & 0.14 &     -1.06 &    0.5 &  1.577 $\pm$ 0.007 &  0.885 $\pm$  0.007 &  7.39 & 1.28\\
      GS4 &  27549 &   53.14471 & -27.78544 &  0.2467 $\pm$ 0.0007 &    7.96 &   5.9 $\pm$ 0.4 &      0.41 $\pm$ 0.04 & 0.83 &     -1.79 &    0.2 &  0.3578 $\pm$ 0.0008 &  0.884 $\pm$  0.006 &  6.82 & 0.71\\
      GS4 &  26639 &   53.14212 & -27.78670 &  0.2270 &      9.02 &   5.9 $\pm$ 0.9 &    0.74 $\pm$ 0.09 & 0.79 &     -0.87 &    1.2 &  0.5704 $\pm$ 0.0003 &  0.5596 $\pm$  0.0009 &  7.05 & 1.39\\
      GS4 &  26696 &   53.19564 & -27.78777 &  0.2270 &      7.86 &   4.6 $\pm$ 0.9 &    0.49 $\pm$ 0.04 & 0.46 &     -1.84 &    0.0 &  1.02 $\pm$ 0.03 &  8.0 $\pm$  0.3 &  7.62 & 0.35\\
      GS4 &  25632 &   53.15464 & -27.79324 &  0.23 $\pm$ 0.01 &        8.26 &   3.9 $\pm$ 0.6 &    0.37 $\pm$ 0.05 & 0.37 &     -1.30 &    0.0 &  0.533 $\pm$ 0.001 &  0.546 $\pm$  0.005 &  7.29 & 0.77\\
      GS4 &  26646 &   53.18153 & -27.78797 &  0.2122 &        7.48 &   1.2 $\pm$ 0.7 &    0.17 $\pm$ 0.02 & 0.17 &     -1.33 &    0.1 &  0.350 $\pm$ 0.002 &  0.52 $\pm$  0.01 &  7.72 & 0.52\\
      GS4 &  27535 &   53.17032 & -27.78526 &  0.277 $\pm$ 0.006 &      7.94 &   0.4 $\pm$ 0.4 &    0.17 $\pm$ 0.02 & 0.17 &     -1.22 &    0.0 &  0.247 $\pm$ 0.002 &  0.96 $\pm$  0.02 &  7.81 & 0.91\\
      GS5 &  43071 &   53.12259 & -27.70791 &  0.16 $\pm$ 0.02 &        8.78 &   6.8 $\pm$ 1.3 &        0.02 $\pm$ 0.02 & 0.02 &     3.89 &    0.2 &  0.746 $\pm$ 0.005 &  1.55 $\pm$  0.02 &  6.99 &1.73\\
 ERSPRIME &  39634 &   53.07824 & -27.72492 &  0.19 $\pm$ 0.02 &        8.54 &   5.7 $\pm$ 1.5 &        0.150 $\pm$ 0.006 & 0.15 &     -1.81 &    0.0 &  0.82 $\pm$ 0.01 &  1.14 $\pm$  0.03 &  7.68 & 1.11\\
 ERSPRIME &  44465 &   53.05034 & -27.70341 &  0.243 $\pm$ 0.005 &      8.35 &   4.0 $\pm$ 1.0 &        0.22 $\pm$ 0.03 & 0.22 &     -0.76 &    0.0 &  0.43 $\pm$ 0.02 &  8.0 $\pm$  0.4 &  6.97 & 1.22\\
\enddata
\tablenotetext{i}{Objects with spectroscopic redshifts are quoted without redshift uncertainty. Objects only with grism redshifts are quoted with uncertainties.}
\tablenotetext{ii}{From the 3D-HST catalog \cite{skel14}}
\tablenotetext{iii}{From the CANDELS/SHARDS catalog \cite{barr19}}
\tablenotetext{iv}{From the GALFIT catalog \cite{van12}}
\end{deluxetable*}
\end{longrotatetable}

\begin{deluxetable*}{llrrr|rrr}[b!]
\tablecaption{TKRS spectroscopy and attenuation-corrected fluxes. \label{tab:TKRS}}
\tablecolumns{8}
\tablenum{2}
\tablewidth{0pt}
\tablehead{
\colhead{Field} &
\colhead{ID} &
\multicolumn{3}{c}{Observed Flux} & \multicolumn{3}{c}{Attenuation-Corrected Flux} \\
& & \multicolumn{3}{c}{$10^{-17}$ erg s$^{-1}$ cm$^{-2}$} & \multicolumn{3}{c}{$10^{-17}$ erg s$^{-1}$ cm$^{-2}$} \\
\cline{3-8} 
\colhead{} & \colhead{} &
\colhead{\Pab\ } & \colhead{\Ha\ } & \colhead{\Hb\ } & \colhead{\Pab\ } & \colhead{\Ha\ } & \colhead{\Hb\ }
}
\startdata
GN1 &  37683 &       3.9 $\pm$1.4 &     34.4 $\pm$1.0 &      8.6 $\pm$0.4 &                 5.4 $\pm$1.9 &               81.8 $\pm$13.0 &               28.6 $\pm$6.4 \\
GN2 &  19221 &      19.4 $\pm$2.1 &    201.3 $\pm$1.7 &     43.0 $\pm$1.0 &                31.8 $\pm$3.6 &              729.0 $\pm$48.7 &              255.0 $\pm$24.1 \\
GN2 &  15610 &      14.5 $\pm$2.3 &     46.0 $\pm$1.1 &     14.2 $\pm$0.7 &                16.4 $\pm$2.8 &               63.6 $\pm$9.5 &               22.2 $\pm$4.7 \\
GN2 &  18157 &       3.9 $\pm$1.6 &     19.3 $\pm$1.6 &      2.7 $\pm$0.8 &                 9.8 $\pm$5.1 &              219.0 $\pm$176.0 &               76.6 $\pm$87.7 \\
GN3 &  34456 &      27.5 $\pm$2.9 &     16.8 $\pm$1.1 &      1.9 $\pm$1.8 &                83.6 $\pm$76.6 &              309.0 $\pm$739.0 &              108.0 $\pm$370.0 \\
GN3 &  34157 &      28.3 $\pm$2.7 &     21.4 $\pm$0.9 &      4.7 $\pm$0.5 &                45.2 $\pm$6.5 &               72.9 $\pm$20.8 &               25.5 $\pm$10.3 \\
GN3 &  33511 &       4.4 $\pm$1.2 &      7.9 $\pm$0.5 &      2.0 $\pm$0.5 &                 6.1 $\pm$2.3 &               18.7 $\pm$12.6 &                6.5 $\pm$6.3 \\
GN3 &  34368 &       4.1 $\pm$1.0 &     33.1 $\pm$1.0 &      7.1 $\pm$0.8 &                 6.8 $\pm$1.8 &              120.0 $\pm$38.5 &               41.9 $\pm$19.2 \\
GN3 &  35455 &       0.8 $\pm$0.6 &     24.0 $\pm$1.5 &      5.5 $\pm$0.7 &                 1.2 $\pm$0.9 &               72.6 $\pm$27.8 &               25.4 $\pm$13.7 \\
GN4 &  24611 &       4.0 $\pm$1.6 &      8.0 $\pm$1.0 &      1.1 $\pm$0.4 &                 9.9 $\pm$5.8 &               85.5 $\pm$93.6 &               29.9 $\pm$46.5 \\
GN5 &  33249 &       3.1 $\pm$0.8 &      4.6 $\pm$0.3 &      2.5 $\pm$0.5 &                 3.1 $\pm$1.0 &                4.6 $\pm$2.7 &                2.5 $\pm$2.1 \\
\enddata

\tablecomments{Attenuation-corrected fluxes are calculated using a \cite{calz00} attenuation model assuming an intrinsic Case B recombination Balmer decrement of $\Ha/\Hb = 2.86$. }
\end{deluxetable*}

\bibliography{library}{}

\end{document}